\documentclass[12pt]{article}

\usepackage{vmargin}   
\setmarginsrb{2cm}{1cm}{2cm}{2cm}{1cm}{1cm}{1cm}{1.6cm}   
\usepackage{amsmath}   
\usepackage{amssymb}  
\usepackage[dvips]{graphicx}  
\usepackage{rotating}   
\usepackage{psfrag} 
\usepackage{fancyhdr}

\author{Karim Noui\thanks{email:noui@phys.univ-tours.fr} \\
Laboratoire de Math\'ematiques et de Physique Th\'eorique \\
F\'ed\'eration Denis Poisson \\
UMR/CNRS 6083, Facult\'e des Sciences et Techniques \\
Parc de Grandmont, 37200, Tours (EU)}

\title{\bf Three dimensional Loop Quantum Gravity: \\
towards a self-gravitating \\
 Quantum Field Theory}
\date{\today} 
 
\begin{document} 
\sloppy
\maketitle

\begin{abstract} 
In a companion paper, we have emphasized the role of the Drinfeld double $DSU(2)$ in the context of three dimensional Riemannian 
Loop Quantum Gravity coupled to massive spinless point particles. We make use of this result to propose a model for a self-gravitating 
quantum field theory (massive spinless non-causal scalar field) in three dimensional Riemannian space. We start by constructing 
the Fock space of the free self-gravitating field: the vacuum is the unique
$DSU(2)$ invariant state, one-particle states correspond to $DSU(2)$ unitary irreducible simple representations and
any multi-particles states is obtained as the symmetrized tensor product between simple representations. The associated quantum field
is defined by the usual requirement of covariance under $DSU(2)$. 
Then, we introduce a $DSU(2)$-invariant self-interacting potential (the obtained model is a Group Field Theory) 
and compute explicitely the lowest order terms 
(in the self-interaction coupling constant $\lambda$) of the propagator and of the three-points function. Finally, we compute the lowest
order quantum gravity corrections (in the Newton constant $G$) to the propagator and to the three-points function.

\end{abstract} 

\newpage

\subsection*{1. Motivations}
The quantization of a self-gravitating field theory is a difficult issue which is still open in Loop Quantum Gravity (LQG).
Nevertheless, this problem is essential and one has to solve it completely and clearly to claim that LQG not only is a candidate 
for pure quantum gravity but also provides a framework to unify the fundamental interactions. Paradoxally, it seems that we have
all the ingredients to find a solution to that problem. On the one hand, LQG (resp. Spin Foam models) offers  
a rigourous framework for a background-independent canonical (resp. covariant or path integral) quantization of general relativity (GR).
On the other hand, QFT is so successful to describe the physics of elementary particles. All the same, there seems to be a large
incompatibility between the two approaches: GR does not fit in QFT and vice versa. It is customary to invoke the
non-renormalizability of GR as the main reason of this disagreement and then usual perturbative QFT techniques are fundamentaly
inadapted to describe a quantum theory of GR. In QFT, the quantization of the self-gravitating field consists in quantizing first the field
in a given (flat or curved) space-time, and then quantizing the metric degrees of freedom by perturbations. 
If this method manifestly fails,
one can try to attack the problem the other way around, i.e. one quantizes gravity before the matter degrees of freedom.

In fact, this idea is an old one but, up to our knowledge, it was only recently that it was realized explicitely by Freidel and Livine
in the framework of three dimensional Riemannian spin-foam models \cite{FreidelLivine}. In fact, they start
with gravity coupled to classical particles, they quantize the gravitational degrees of freedon through the Ponzano-Regge spin-foam model
and after then they quantize the matter field degrees of freedom allowing for creations and annihilations of point particles. 
Thus, they have shown that 3D quantum gravity
amplitude, in the context of the Ponzano-Regge model coupled to point particles \cite{FreidelLouapre}, are actually the Feynman diagram
evaluation of a braided non-commutative quantum field theory. This effective field theory describes the dynamics of a scalar field
after integrating out the gravitational degrees of freedom. 

This article is devoted to analyse this result in the hamiltonian point of view. For that purpose, we use the results presented in a 
recent companion paper \cite{Noui} which states that the Drinfeld double $DSU(2)$ appears to be 
the ``quantum symmetry group'' of three dimensional
riemannian loop quantum gravity coupled to massive spinless point particles once one imposes the hamiltonian constraint.
To be more precise, we have shown that any multi-particles 
physical states are defined by a tensor product of simple unitary irreducible representations (UIR) of $DSU(2)$ and 
the physical scalar product
between two such states is given in term of the symmetric (or Barrett-Crane) interwiners (see \cite{FNR} for a general definition
and examples of the symmetric intertwiner). Therefore, we start from the simple UIR of the Drinfeld double and construct a Fock space as
the direct sum of the symmetrized tensor product of these representations. Then, we define creation and annihilation operators
acting on this Fock space and we construct a local self-gravitating quantum field $\phi$ by the usual requirement that it transforms
covariantly under the Drinfeld double transformations. This defines the model we propose for a free self-gravitating scalar non-causal
quantum field theory. 
By non-causal, we mean that Feynmann graphs amplitudes are those of a non-causal spin-foam models: therefore, the model admits no
causal structure and no dynamics. In a sense, it behaves as a topological quantum scalar field theory.
Finally, we introdu
ce a (cubic) self-interaction in the model and we show the effects of quantum gravity on some physical processes
like particles propagation or particles creation. 
In particular, the propagator of the self-gravitating free field, given by the two-points function $\Delta_G(x,y)$
($G$ being the Newton constant), is computed
and is manifestly different from the non-gravitating massive scalar field one $\Delta_0(x,y)$: 
we show that there are quantum gravity corrections to
the classical propagator which are in total agreement with those obtained in the context of spin-foam models by Freidel and Livine. 
In the limit where $G$ tends to zero, the self-gravitating propagator $\Delta_G(x,y)$ tends to the non-gravitating
propagator $\Delta_0(x,y)$. Then, we concentrate on the three points function of the self-interacting theory: we compute lowest order 
terms (in the self-interaction coupling constant) in both self-gravitating and non-gravitating cases; thus
we are able to compute explicitely the quantum gravity effects on the three points function.

\medskip

The paper is organized as follows. The section 2 is devoted to the construction (\`a la Wigner as recalled in Weinberg book) 
of the self-gravitating quantum field. We start by briefly recalling the construction in the case of a scalar field in three
dimensional euclidean space: definitions of one-particle states, of the Fock space and of creation and annihilation operators.
Then, we adapt this construction to define our model for a self-gravitating non-causal 
quantum field theory on a sphere: we define one-particle and
multi-particles states in term of particles-spin-networks, we construct naturally the Fock space and define the self-gravitating quantum
field as on operator that transforms covariantly under $DSU(2)$ transformations. We show that our model defines in fact a non-commutative
braided quantum field theory which is in total agreement with the results obtained by Freidel and Livine in the spin-foam context.
We compute the self-gravitating propagator (two-points function) and exhibit the first quantum gravity corrections to the classical 
propagator.

In the section 3, we give the lagrangian formulation of our model. Then, we introduce a self-interaction  (cubic term in the field) 
and compute lowest order (in the self-interaction coupling constant $\lambda$) in terms of the self-gravitating three-points function. 

To conclude, we discuss the possibility to generalize our construction to the lorentzian case as well as to the case where 
there is a non-vanishing positive or negative cosmological constant.

\subsection*{2. A self-gravitating massive Quantum Field Theory}
This section aims at presenting the construction of the self-gravitating quantum field theory in the canonical framework.
Following Weinberg approach \cite{Weinberg} based on Wigner analysis, 
we first recall the construction of a bosonic scalar field in three dimensional euclidean
space. In a second time, we adapt this method to construct and study the self-gravitating quantum scalar field.

\subsubsection*{2.1. A simple example: the scalar field in three dimensional euclidean space}
We start by recalling the construction of a massive quantum field in the three dimensional euclidean space $\mathbb E^3$ ($\mathbb E^3$
is the space $\mathbb R^3$ endowed with the euclidean metric $\text{diag}(+,+,+)$). 

\medskip

The unitary irreducible representations of the universal covering of the isometry group of $E^3$, denoted $ISU(2)=\mathbb R^3 \times 
SU(2)$, 
provide the one-particle space of states of the quantum field. These representations are classified by a mass $m \in \mathbb R$ 
and a spin $s \in \frac{1}{2} \mathbb N$. Among these representations, one distinguishes the simple ones,
caracterized by the fact that $s=0$,
whose associated vector space ${\cal H}$ is the Hilbert space of states for the spinless massive particle on $\mathbb E^3$.
The one particle Hilbert space is ${\cal H}[m] \simeq (L^2(S^2),d\mu)$ where $L^2(S^2)$ is the set of functions on 
the two-sphere $S^2=SU(2)/U(1)$ which are squared integrable with respect to the normalized measure $d\mu$. In the sequel,
we will identify a point $\lambda \in S^2$ with an element $\lambda \in SU(2)$ that is a representative of a given 
conjugacy class of $[\lambda] \in SU(2)/U(1)$.

The action of any element $(\vec{x},g) \in \mathbb R^3 \times SU(2)$ on states $\varphi \in {\cal H}[m]$ associated to
a particle of mass $m$ reads:
\begin{eqnarray}\label{ISU(2)representation}
(\pi_m(\vec{x},g) \; \varphi)(\lambda) \; = \; \exp(i \vec{x} \cdot \lambda \vec{m}) \;  \varphi(g^{-1} \lambda)
\end{eqnarray}
where $\lambda \vec{m}$ denotes the action of $\lambda \in SU(2)$ on the ``rest'' vector $\vec{m}=(m,0,0) \in \mathbb R^3$. 
It is customary to extend the space of states to the set of distributions on $S^2$ in order to include pure 
momentum states into ${\cal H}[m]$. In that case, one introduces the usual bra-ket notation $\vert \lambda >$ 
 and the representation (\ref{ISU(2)representation}) is trivially rewritten as
follows:
\begin{eqnarray}
\pi_m(\vec{x},g) \; \vert \lambda > \; = \; \exp(i \vec{x} \cdot \lambda \vec{m}) \; \vert g^{-1} \lambda>
\end{eqnarray}
In the sequel, we will mostly use this notation which is more familiar in Quantum Field Theory. Note that the duality bracket is
given as usual by $<\lambda\vert\varphi>=\varphi(\lambda)$.

To construct a bosonic field from these representations, one starts by defining the notion of a $n$ particles state represented by
elements $\vert \lambda_1,\cdots,\lambda_n>$ that belong to the symmetrized tensor product of one-particle state, i.e.:
\begin{eqnarray}
\vert \lambda_1,\cdots,\lambda_n> \; = \; \frac{1}{n!}
\sum_{\sigma \in P_n} \vert \lambda_{\sigma(1)}>\otimes \vert \lambda_{\sigma(2)}> \otimes 
\cdots \otimes \vert \lambda_{\sigma(n)}>
\end{eqnarray}
where the sum runs over the permutation group $P_n$. Note that there is no analog of the well-known four dimensional spin-statistic 
theorem for three dimensional quantum fields theories (see \cite{SpinStatistic} for example). In particular, there exist 
in three dimensional space-time exotic statistics which interpolate between bosonic and fermionic statistics (as it is briefly
explained in the appendix A). 
Nevertheless, a spinless quantum field in flat euclidean space can only be bosonic if one requires locality, covariance and causality;
there is no restriction on the statistics if one considers a spinning quantum field as explained in the appendix A.
As we deal with spinless massive field, we will only consider the bosonic statistic here.
In that case, the space of $n$-particles states is given by ${\cal H}^s_n[m] = {\cal H}[m]^{\otimes_s n}$ 
where $\otimes_s$ stands for the symmetrized tensor product. 
The bosonic Fock space ${\cal F}[m]$ for a massive field of mass $m$ is then defined as the tower of multi-particles states, i.e.:
\begin{eqnarray}
{\cal F}[m] \; \equiv \; \bigoplus_{n=0}^{\infty} {\cal H}^s_n[m] \; = \; \bigoplus_{n=0}^{\infty} {\cal H}[m]^{\otimes_s n}\;.
\end{eqnarray}
The space ${\cal H}_0[m] \simeq {\cal H}_0[0]$ is the trivial representation Hilbert space of $ISU(2)$. The Fock space carries in
fact a reducible representation of $ISU(2)$ whose action on ${\cal F}[m]$ simply reads:
\begin{eqnarray}\label{actionISU(2)}
ISU(2) \times {\cal F}[G] & \longrightarrow & {\cal F}[G] \nonumber \\
(\vec{x},g) \times \vert \lambda_1,\cdots,\lambda_n>  & \longmapsto & U(\vec{x},g) \; \vert \lambda_1,\cdots,\lambda_n> \; \equiv \;
\pi_m^{\otimes n} \Delta^{(n)}(\vec{x},g) \; \vert \lambda_1,\cdots,\lambda_n> \;,
\end{eqnarray}
where $\Delta^{(n)} : ISU(2) \rightarrow ISU(2)^{\otimes n}$ is the iterated co-product defined by:
\begin{eqnarray}
\Delta^{(1)}(\vec{x},g)=\Delta(\vec{x},g)=(\vec{x},g) \otimes (\vec{x},g) \;\;\;\; \text{and} \;\; 
\Delta^{(n+1)} = id \otimes \Delta^{(n)}.
\end{eqnarray}
Note that $\Delta$ is the usual co-commutative coproduct defined for groups. 

Bosonic creation and annihilation operators are denoted as usual $a^\dagger(\lambda)$ and $a(\lambda)$, satisfy
the following commutation relations:
\begin{eqnarray}\label{commutationrelations}
[a^\dagger(\lambda_1),a^\dagger(\lambda_2)] \; = \; 0 \; = \; [a(\lambda_1),a(\lambda_2)] \;\;\; \text{and} \;\;\;
[a(\lambda_1),a^\dagger(\lambda_2)] \; = \; \delta(\lambda_1^{-1}\lambda_2)\;, 
\end{eqnarray}
and acts on the Fock space by respectively raising or lowering the number of particles. There is a natural co-action of $ISU(2)$ 
on the set of creation and annihilation operators defined by duality from the action (\ref{actionISU(2)}). 
It is customary to describe this co-action with the following notations:
\begin{eqnarray}\label{coactionISU(2)}
a_{\pm}(\lambda) \; \longmapsto \; U(\vec{x},g) a_{\pm}(\lambda) U(\vec{x},g)^{-1} \; = \;  
\exp (\mp i \vec{x} \cdot \lambda \vec{m}) \; a_{\pm}(g^{-1}\lambda) 
\end{eqnarray}
where $a_+(\lambda) = a^\dagger(\lambda)$ and $a_-(\lambda) = a(\lambda)$ are the creation and annihilation operators.

\medskip

We have now all the pieces to construct a free local quantum field. Following Weinberg \cite{Weinberg}, the quantum field
$\phi$ is defined as a Fock space operator valued function on the euclidean space $\mathbb E^3$ that satisfies the fundamental
properties of locality, covariance and causality. Locality is automatically satisfied as $\phi$ is a function, i.e. is defined for
each point $\vec{x}$ of $\mathbb E^3$ as follows:
\begin{eqnarray}
\phi(\vec{x}) \; \equiv \; \int d\mu(\lambda) \; (c_+(\lambda,\vec{x}) a^\dagger(\lambda) \; + \; c_-(\lambda,\vec{x}) a(\lambda))\;.
\end{eqnarray}
The covariance means that the field transforms in the same way as creation and annihilation
operators under the action of the elements of $ISU(2)$ (\ref{coactionISU(2)}). As an immediate consequence, the quantum field is 
completely determined by its value at the origin $\phi(\vec{0})$ which is co-invariant under the action of the rotational 
subgroup $SU(2) \subset ISU(2)$. It follows that:
\begin{eqnarray}
U(\vec{0},g) \phi(\vec{0}) U(\vec{0},g)^{-1} \; = \; \phi(\vec{0}) \;\;\;\; \Longrightarrow \;\;\;\; c_{\pm}(\lambda,\vec{x}) \; = \;
A_{\pm} \; \exp(\mp i \vec{x} \cdot \lambda \vec{m})  
\end{eqnarray}
where $A_{\pm}$ can be choosen to be real numbers. Causality is meanningful only when the underlying
base space admits a causal structure: in that case, a field is said causal if $[\phi(x),\phi(y)]=0=[\phi(x)^\dagger,\phi(y)]$ 
when $x$ and $y$ are causally disconnected points. In our case, $\mathbb E^3$ is an euclidean space and we replace the causality property
by the requirement that $[\phi(x),\phi(y)]=0=[\phi(x)^\dagger,\phi(y)]$ 
for each couple of points $(x,y)$. As a consequence, $A_+=A_-=A$ and the quantum field is finally given by:
\begin{eqnarray}
\phi(\vec{x}) \; = \; A \int d\mu(\lambda) \; (\exp(-i \vec{x} \cdot \lambda \vec{m}) a^{\dagger}(\lambda) + 
\exp(i \vec{x} \cdot \lambda \vec{m}) a(\lambda)) \;.
\end{eqnarray}
Finally, the propagator of this free theory is given by the two-points function which reads:
\begin{eqnarray}\label{ISU(2)2points}
\Delta_0(\vec{x},\vec{y}) \; \equiv \; <0\vert \phi(\vec{x}) \; \phi(\vec{y}) \vert 0> \; = \; A^2 \;
\frac{\sin m \vert \! \vert \vec{x} - \vec{y} \vert \! \vert}{m \vert \! \vert \vec{x} - \vec{y} \vert \! \vert} \;.
\end{eqnarray}
The propagator is defined here up to a global constant\footnote{The change of variable 
$\vec{P}=\lambda \vec{m}$ and
$d\mu(\lambda)= \frac{1}{2\pi m}\delta(\vec{P} \cdot \vec{P} - m^2) d^3\vec{P}$
in the previous expression
allows to reexpress the field $\phi$ in the more familiar form involving the full momentum $\vec{P}$:
\begin{eqnarray*}
\phi(\vec{x}) \; = \; \tilde{A} \int d^3\vec{P} \; \delta(\vec{P} \cdot \vec{P} - m^2) \;
(\exp(-i \vec{x} \cdot \vec{P}) a^{\dagger}(\vec{P}) + \exp(i \vec{x} \cdot \vec{P}) a(\vec{P})) \;.
\end{eqnarray*}
where $\tilde{A}=A/(2\pi m)$ and $a(\lambda) \equiv a(\lambda \vec{m}) \equiv a(\vec{P})$.
Usually in QFT, one writes the quantum field as an integral involving only the space momentum $\vec{p}$
and not the ``space-time'' momentum $\vec{P}$. To do so, one integrates over the time component $P_0$ and one imposes a causal structure
by the constraint $P_0>0$. However, not only there is no canonical way to exhibit a time component $\vec{P}_0$ out of an 
euclidean momentum $\vec{P}$ but also there is no consistent way to impose the constraint ${P}_0>0$ for the space of euclidean 
momentum of fixed mass $m$ is a sphere whereas it is a double-connected hyperboloid in the Lorentzian case (what makes to positivity 
condition valuable). If we all the same naively impose a causal structure (with an Heaviside function $\Theta(P_0)$), 
the scalar field would be defined as follows:
\begin{eqnarray*}
\phi(\vec{x}) \; = \; \tilde{A} \int d^3\vec{P} \; \delta(\vec{P} \cdot \vec{P} - m^2) \; \Theta (P_0) \;
(\exp(-i \vec{x} \cdot \vec{P}) a^{\dagger}(\vec{P}) + \exp(i \vec{x} \cdot \vec{P}) a(\vec{P}))\;.
\end{eqnarray*}
One could integrate over the variable $\vec{P}_0$ to have an expression in terms of $\vec{p}$. Finally,
the equal time canonical relation
$
[\phi(x_0,\vec{x}_s),\partial_{x_0} \phi(x_0,\vec{y}_s)] \; = \; i \; \delta^{(2)}(\vec{x}_s - y_{s})
$
fixes the value of $\tilde{A}$. The Planck constant if fixed to $\hbar=1$.} $A$. In the following, we will take the value $A=1$.

Note that the propagator is in fact the Hadamard function of the field: 
it is a solution of the Klein-Gordon equation but not a Green function. 
To recover the usual Feynman propagator, one has to choose a causal structure which means making a choice of a time variable and then
defining the two-points function (\ref{ISU(2)2points}) as the expectation value of the chronological product of fields. There is a
natural way to do so if we deal with a lorentzian quantum field theory instead of an euclidean one.

\subsubsection*{2.2. A Quantum scalar field on a quantum euclidean background}
The idea is now to adapt this well-understood method to construct a self-gravitating three dimensional massive quantum field theory.
We will first present the construction of the Fock space of the theory starting with the description of the one-particle state
and the multi-particles states. Then, we will construct the quantum fields requiring the basic properties of locality, covariance
and causality. As we deal with hamiltonian quantization, we will exclusively consider the case where the topology of three dimensional 
space-time ${\cal M}$ is ${\cal M} = \Sigma \times [t_1,t_2]$. Moreover, we concentrate on the spherical case only, i.e. $\Sigma=S^2$.

\vspace{0.3cm}

{\it a. One-particle states.}

\medskip

A system of gravitating particles is defined with respect to one observer ${\cal O}$ whose mass is fixed to the value $m_0$.
In fact, $m_0$ is the total mass of the system of particles. In the gravitational case, states 
of a spinless massive particle in the sphere are elements of ${\cal H}[m] \times {\cal L}(1)$ where ${\cal H}[m]=(L^2(S^2),d\mu)$
is the Hilbert space of one-particle state in $\mathbb E^3$ and ${\cal L}(1)$ denotes the set of oriented links between one particle 
and the observer. 
As previously, we will prefer for clarity reasons to extend ${\cal H}[m]$ to distributional states and introduce $\vert \lambda>$ that
represents a pure momentum state written in the bra-ket notation.
Thus, a one-particle state for a self-gravitating quantum field is
denoted $\vert \lambda,\ell>$. The spin-network evaluation of this state of quantum gravity is rigourously given by:
$<\varphi,A \vert \lambda,\ell> = \varphi (H_\ell(A) \lambda)$ where $A$ is a $SU(2)$ flat connection on $\Sigma$, 
$\varphi \in {\cal H}[m]$ caracterises the particle state and $H_\ell(A)$ is the holonomy of the connection along the oriented 
link $\ell$. As the gravitational degrees of freedom are pure gauge, we always omit to mention the flat connection that is implicitely
gauge-fixed to the trivial one.  
Because of diffeomorphisms invariance and the trivial topology of $S^2$, 
$\vert \lambda, \ell>$ and $\vert \lambda, \ell'>$ are physically equivalent whatever the links $\ell$ and $\ell'$ are. 

\vspace{0.3cm}

{\it b. Multi-particles states: symmetrization vs. diffeomorphisms invariance.}

\medskip

\begin{figure}[h]
\psfrag{1}{$1$}
\psfrag{2}{$2$}
\psfrag{3}{$3$}
\psfrag{n}{$n$}
\psfrag{...}{$\cdots$}
\psfrag{l1}{$\ell_1$}
\psfrag{l2}{$\ell_2$}
\psfrag{l3}{$\ell_3$}
\psfrag{ln}{$\ell_n$}
\psfrag{lp1}{$\ell'_1$}
\psfrag{lp2}{$\ell'_2$}
\centering
\includegraphics[scale=0.8]{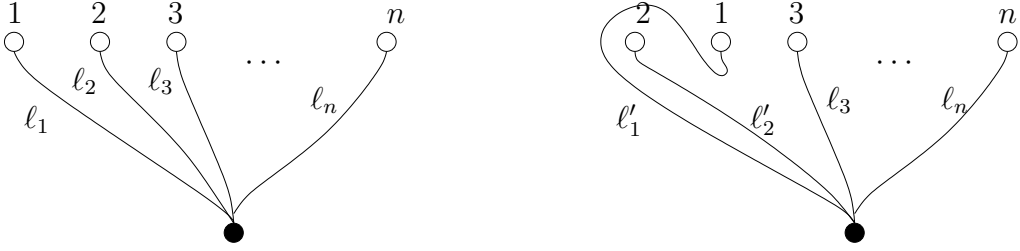}
\caption{Pictorial representation of a multi-particles state coupled to quantum gravity. The two states are in fact physically 
equivalent for they are related by a spacial diffeomorphism.}
\label{multiparticles}
\end{figure}

Let us now describe multi-particles states. Any $n$-particles state on the surface $\Sigma$
is caracterized by a set of $n$ points or circles $(x_1,\cdots,x_n)$ on $\Sigma$ corresponding to the ``locations'' of the particles,
a family of momenta $(\lambda_1,\cdots,\lambda_n)$ associated to each particle and a family of oriented links
$(\ell_1,\cdots,\ell_n)$ between the observer ${\cal O}$ and each particle (see figure \ref{multiparticles}). The links are oriented
from the observer to the particles. 

In fact, the set of links forms a minimal graph $\gamma_n$ which defines a $n$-particles-spin-network state, i.e. a quantum state of 
the coupled system $\{\text{gravity}+\text{particles}\}$. This structure has been introduced and studied in \cite{NouiPerez, Noui}.
Such a state is represented as a tensor product of one-particle states. 
The Hilbert space of $n$-particles states is a sub-space of 
${\cal H}_n[m] \equiv {\cal H}[m]^{\otimes n} \times {\cal L}(n)$ where ${\cal L}(n)$ 
denotes the set of links between $n$ particles and the observer. It will be convenient in the sequel to adopt the following notation
for elements of ${\cal H}_n[m]$:
\begin{eqnarray}\label{Gmultistates}
\vert \lambda_1,\ell_1> \otimes \cdots \otimes \vert \lambda_n,\ell_n> \; \equiv \; \vert \lambda_1 \otimes \cdots \otimes \lambda_n;
\ell_1 \otimes \cdots \otimes \ell_n> \; \equiv \; \vert \otimes_i \lambda_i;\gamma_n> \;.
\end{eqnarray}
This notation makes a clear distinction between the ${\cal H}[m]^{\otimes n}$ part from the ${\cal L}[n]$ part in ${\cal H}_n[m]$
and means that: the point $x_i$ ($i^{th}$ element on the tensor product) is associated to a particle of momentum $\lambda_i$ and is
linked to the observer by $\ell_i$. By convention, $\ell_i \otimes \ell_j$ means that $\ell_i < \ell_j$  for the order on the set
of links on a given graph $\gamma_n$ ;
it will be convenient for what follows to introduce the notation $\ell_1 \otimes^{op} \ell_2$ for a given graph $\gamma_2$ which 
means that $\ell_2<\ell_1$.
We will denote ${\cal H}_n[m,\gamma_n]$ the space of states defined on a given graph $\gamma_n$.

To evaluate such a spin-network, we fix a function $\varphi \in {\cal H}[m]^{\otimes n}$, a $SU(2)$ connection $A$ on $\Sigma$ and
we have:
\begin{eqnarray}
<\varphi, A \vert \otimes_i \lambda_i; \gamma_n> \; = \; 
\varphi(H_{\ell_1}(A) \lambda_1 \otimes \cdots \otimes H_{\ell_n}(A) \lambda_n) \;.
\end{eqnarray}
We have implicitely assumed that $\ell_1 < \cdots < \ell_n$ in the graph $\gamma_n$ (otherwise we have to perform a permutation in the 
evaluation of the spin-network).
As for one-particle states, we omit the gravitational part $H_\ell(A)$ by gauge fixing the connection to the trivial one.
This gauge fixing does not affect, by definition, physical observations and predictions (see \cite{Noui} for details).

What are the properties of multi-particles state under permutations of particles?
To answer this question, we first have to precise what we mean by permuting two particles as we needed three different orders
to define ${\cal H}_n[m,\gamma_n]$. Note that the observer is sensitive to the order on the links: from his point of view, 
the particles are ordered according to the order of the links
in the sense that a particle 1 is on the left of a particle 2 if the link $\ell_1 < \ell_2$ (i.e. $\ell_1$ is on the left
of $\ell_2$ at the level of the observer). 
Usually, we choose the orders on the particles to be the one inherited from the order on the links 
(case of the left picture of the figure \ref{multiparticles}). We make this choice in the following.
A permutation of the multi-particles state is defined by permuting the momenta of the particles leaving unchanged the order of 
the points, of the links and the shape of the minimal graph $\gamma_n$,
i.e.:
\begin{eqnarray}\label{permutation}
\sigma \in P_n \; : \; \vert \otimes_i \lambda_i;\gamma_n> \; \mapsto \; \vert \sigma(\otimes_i \lambda_i);\gamma_n>
\end{eqnarray}
where $\sigma$ is an element of the permutation group $P_n$ of $n$ elements. 

As we want to describe a bosonic quantum field, we require
that two states (\ref{Gmultistates}) that differs only by a permutation (\ref{permutation}) are physically indistinguishable. 
Therefore, the definition of a $n$-particles physical state for the bosonic quantum field is the following one:
\begin{eqnarray}
\vert \otimes_i^s \lambda_i;\gamma_n> \equiv 
\frac{1}{n!} \sum_{\sigma \in P_n} \vert \sigma (\otimes_i \lambda_i); \gamma_n> \;.
\end{eqnarray}
The space of a bosonic self-gravitating $n$-particles states is denoted ${\cal H}^s_{n}[m;\gamma_n]$
and depends a priori on the given minimal graph $\gamma_n$. 
We choose $\gamma_{n+1} \supset \gamma_n$ such that there exists a canonical inclusion 
${\cal H}^s_n[m;\gamma_n] \hookrightarrow {\cal H}^s_{n+1}[m;\gamma_{n+1}]$. Finally, the bosonic Fock space 
for a self-gravitating quantum field of mass $m$ is given as usual by the following infinite direct sum:
\begin{eqnarray}\label{Fockspace}
{\cal F}[m;\gamma] \; \equiv \; \bigoplus_{n=0}^{\infty} {\cal H}^s_{n}[m;\gamma_n] \;
\end{eqnarray}
where we have introduced the notation $\gamma = \otimes_{i=1}^{\infty} \ell_i = \cup_{n=1}^\infty \gamma_n$. 
The structure of the Fock space for the self-gravitating quantum field is similar to those of the space of cylindrical functions 
on $\Sigma$ defined by means of projective limits. 
As we will show in
the sequel, the graph $\gamma$ is not physically relevant in the definition of the Fock space: once we construct the quantum field,
we will see that physical quantities as $n$-points functions does not depend on the choice of $\gamma$. This is directly linked
to diffeomorphisms invariance of quantum gravity for any two graphs $\gamma$ and $\gamma'$ on the sphere $\Sigma$ are related by
a spacial diffeomorphim.

Let see what happens in the simple example where one turns one particle around another one without changing the homotopy class
of the graph (figure \ref{multiparticles}). The two states belongs to different Fock spaces but should be physically equivalent.
This particles transposition is described in terms of a map $\tau:{\cal H}_2[m,\gamma_2] \rightarrow {\cal H}_2[m,\gamma'_2]$ 
(its inverse $\tau^{-1}$) given by:
\begin{eqnarray}
\tau: \vert \lambda_1 \otimes \lambda_2; \ell_1 \otimes \ell_2>
& \longmapsto & 
\vert \lambda_2 \otimes \lambda_1; \ell_1 \otimes^{op} \ell_1 c_1^- \ell_1^{-1}\ell_2 > \label{tau}\\
\tau^{-1}: \vert \lambda_1 \otimes \lambda_2; \ell'_1 \otimes \ell'_2> & \longmapsto & 
\vert \lambda_2 \otimes \lambda_1;\ell'_2 c_2^+ \ell'_2{}^{-1} \ell'_1 \otimes^{op} \ell'_1> \label{tau-}
\end{eqnarray}
where the product between the set of links is the usual composition and $c_i^+$ (resp. $c_i^-$) is the anti-clockwise
(resp. clockwise) loop around the particle located at the point $x_i$. 
These maps are trivially extended to the Fock space.
For clarity reasons, we restrict ourselves to the case $n=2$: 
using notations of the picture (\ref{multiparticles}), we have $\gamma_2=\ell_1 \otimes \ell_2$ and 
$\gamma'_2=\ell'_1 \otimes \ell'_2$ with $\ell_1'=\ell_1 c_1^- \ell_1^{-1} \ell_2$ and $\ell_2'=\ell_1$.
Note that generically $\tau^2 \neq 1$ and therefore there are infinitely many denombrable
inequivalent ways to transpose two particles: this is a well-known particularity of 2+1 dimensional quantum field theory. 

In fact, the relations (\ref{tau},\ref{tau-}) define a representation of the braid group: this representation is not one-dimensional
and is isomorphic to a representation of the $R$-matrix of the Drinfeld double $DSU(2)$. To understand this point precisely, we evaluate
the following spin-networks:
\begin{eqnarray*}
<\varphi,A \vert \tau \vert \lambda_1 \otimes \lambda_2; \ell_1 \otimes \ell_2> & = & 
\varphi( H_{\ell_1}(A)\lambda_2 \otimes H_{\ell_1}(A) \lambda_2 h(m)^{-1}\lambda_2^{-1} H_{\ell_1}^{-1}(A) H_{\ell_2}(A) \lambda_1) \\
<\varphi,A \vert \tau^{-1} \vert \lambda_1 \otimes \lambda_2; \ell_1 \otimes \ell_2> & = & 
\varphi(H_{\ell_2}(A) \lambda_1 h(m)\lambda_1^{-1} H_{\ell_2}^{-1}(A) H_{\ell_1}(A) \lambda_2 \otimes H_{\ell_2}(A)\lambda_1)
\end{eqnarray*}
where $\varphi \in {\cal H}_2[m]$, $A$ a $SU(2)$ flat connection and $H_{c_i^{\pm}}=\lambda_i h(m)^{\pm 1} \lambda_i^{-1}$ 
with $h(m) \in SU(2)$ such that $h(m)=\text{diag}(e^{im},e^{-im})$ in the fundamental representation. Once we gauge fixed the connection
to the trivial one ($A=0$), we immediately see that:
\begin{eqnarray}\label{diffbraid}
<\varphi,0 \vert \tau^{\varepsilon} \vert \lambda_1 \otimes \lambda_2; \ell_1 \otimes \ell_2> & = & 
<(\pi_m \otimes \pi_m)R^{\varepsilon}\varphi,0 \vert \lambda_1 \otimes \lambda_2; \ell_1 \otimes \ell_2>
\end{eqnarray}
where $\varepsilon \in \{+1,-1\}$, $R$ (resp. $R^{-1}$) is the $DSU(2)$ (resp. inverse) R-matrix
and $\pi_m$ denotes the simple representation labelled by the mass $m$ of the field. Thus, we have a clear relationship
between particles transpositions and $DSU(2)$ braidings (see appendix B for details). 
This example illustrates the close relationship between $DSU(2)$ braidings and diffeomorphisms. It can be easily generalized.

We finish this section with some important remarks. First, we have to show at the end of our construction that physical quantities
like $n$-points functions are invariant under braidings. Second, it is clear that multi-particles have to be viewed as representations
of $DSU(2)$ instead of representations of $ISU(2)$. As Hilbert spaces, representations of $DSU(2)$ and $ISU(2)$ are isometric.
But the action of the group operators on the states will be different and that makes the essentiel difference in the construction
of the self-gravitating quantum field compared to the usual quantum field in $\mathbb E^3$.

\vspace{0.3cm}

{\it c. Gravitational deformation of translations and non-commutative space-time.}

\medskip

The Drinfeld double $DSU(2)$ is clearly the quantum symmetry group of a system of point massive particles coupled to
Riemannian three dimensional quantum gravity. This has been shown in the context of combinatorial quantization \cite{BKM},
of Ponzano-Regge spin-foam models \cite{FreidelLouapre} and recently in the context of Loop Quantum Gravity \cite{Noui}. 
In fact, $DSU(2)$ is a gravitational deformation of $ISU(2)$ (appendix B) and can be viewed as the symmetry group of the
system of self-gravitating particles once quantum gravity effects have been taken into account. Thus, quantizing a system of 
self-gravitating massive particles on $\mathbb E^3$ is equivalent to quantizing a system of non-gravitating massive point particles
whose symmetry group is $DSU(2)$ instead of $ISU(2)$. As a consequence, $n$-particles states transform as representations of $DSU(2)$
under rotations and translations; and these transformations laws are deformed compared to usual $ISU(2)$ ones. Finally, the massive
self-gravitating quantum field will naturally be defined by the requirement that it transforms covariantly under $DSU(2)$.

Before going to the construction of the field operator, we present basic properties concerning the Drinfeld double.
Following notations of the appendix B, a (distributional) element of the Drinfeld double is given by a pair $(g,u)$ of $SU(2)$
elements. In this notation, the product and co-product of $DSU(2)$ read:
\begin{eqnarray}
(g_1,u_1) \cdot (g_2,u_2) & = & \delta(g_1^{-1}u_1g_2u_1^{-1})(g_1,u_1u_2) \label{DSU(2)grouplaw}\\
\Delta(g,u) & = & \int dh \; (gh,u)\otimes(h^{-1},u)\;. \label{DSU(2)co}
\end{eqnarray}
Note that $\delta$ is the delta distribution on the group $SU(2)$ and $\int dh$ is the $SU(2)$ Haar measure.
In the sequel, we will also need the expression of the antipode: $S(g,u)=(u^{-1}g^{-1}u,u^{-1})$. 
Comparing this structure with $ISU(2)$ one, it is clear that $DSU(2)$ is a deformation of $ISU(2)$. In fact, 
the rotational part remains the same as for $ISU(2)$ and is isomorphic as a Hopf algebra to $SU(2)$
whereas the translation part structure is deformed and is no longer isomorphic to $\mathbb R^3$. 

The deformation of $ISU(2)$ into $DSU(2)$ is caracterized by a group algebras morphism $\phi: ISU(2) \rightarrow DSU(2)$ 
which is obviously not a co-algebra morphism (appendix B). The map shows that $g$ has to be interpreted as a deformation of a momentum
on $\mathbb E^3$ and not strickly as a deformation of a position in $\mathbb E^3$. In order to recover gravitational deformed analogs
of the positions variables, one has to introduce the following deformed Fourier transform:
\begin{eqnarray}\label{qFT}
(\vec{x},u) \; \equiv \; \int dg \; e^{i \text{tr}(gx)}\; (g,u) \; 
\end{eqnarray}
where we make used of the identification $\mathbb E^3 \rightarrow su(2); \vec{x} \mapsto x=\vec{x}\cdot \vec{\sigma}$ with
$\vec{\sigma}=(\sigma_0,\sigma_1,\sigma_2)$ the generators of the Lie algebra $su(2)$; 
$\text{tr}$ denotes the trace in the fundamental representation.  The group law satisfied by the elements $(\vec{x},u)$ is trivially 
obtained from the relation (\ref{DSU(2)grouplaw})
\begin{eqnarray}
(\vec{x}_1,u_1) \cdot (\vec{x}_2,u_2) \; = \; (\vec{x}_1 + u_1 \vec{x}_2,u_1u_2)
\end{eqnarray}
and it becomes clear that $\vec{x}$ is the very analog of the position. The difference with positions variables in $\mathbb E^3$ is that
the variables (\ref{qFT}) are in fact position variables on a non-commutative space-time. 

Let us be more precise. First, we introduce the notation $\mathbb E^3_G = DSU(2)/SU(2)$. $\mathbb E^3_G$ admits a Hopf algebra structure.
The space of functions on $\mathbb E^3_G$ denoted $Fun(E^3_G)$ inherits by duality a Hopf-algebra structure. 
Of particular interest are the plane waves on $\mathbb E^3_G$ defined for any $g\in SU(2)$ by 
\begin{eqnarray}
w_g \;: \; \mathbb E^3_G \; \longrightarrow \; \mathbb C \;\;\;\;\;  \vec{x} \; \mapsto \; w_g(\vec{x}) \;= \; e^{i\text{tr}(xg)} 
\end{eqnarray}
where we identify $\mathbb E^3_G$ and $su(2)$ as in (\ref{qFT}). In terms of these deformed plane waves, the product
and co-product $\Delta_F$ of $Fun(\mathbb E^3_G)$ respectively read:
\begin{eqnarray}
(w_g \; \star \; w_h)(\vec{x}) & \equiv &  (w_g \otimes w_h)(\Delta(\vec{x})) \; = \; w_{gh}(\vec{x}) \\
\Delta_F(w_g)(\vec{x} \otimes \vec{y}) & \equiv & w_g ((\vec{x},1) \cdot (\vec{y},1)) \; = \; w_g(\vec{x}+\vec{y})\;.
\end{eqnarray}
Due to the non-commutativity of $SU(2)$, the product $\star$ on $Fun(\mathbb E^3_G)$ is non-commutative and therefore the space 
$\mathbb E^3_G$ is a non-commutative space-time. In the non-gravitational limit, $\mathbb E^3_G$ tends to the usual euclidean spacetime
$\mathbb E^3$ and therefore becomes commutative as expected. Note that the non-commutativity is a direct consequence of the 
non-co-commutativity of the Drinfeld double. The product on $Fun(\mathbb E^3_G)$ is closely related to the convolution
product $\circ$ on the space of functions on $SU(2)$ denoted $Fun(SU(2))$. Indeed, the following map: 
\begin{eqnarray}
(Fun(SU(2)),\circ) \; \rightarrow \; (Fun(\mathbb E^3_G),\star) \;\;\;\;\; \tilde{f} \; \mapsto \; f:\vec{x} \mapsto 
f(\vec{x}) \; = \; \int dg \; \tilde{f}(g) \; e^{i\text{tr}(gx)} 
\end{eqnarray}
is an algebra morphism. To be more precise, the map is a surjection whose kernel $Ker \simeq \mathbb Z_2$ is given by the set  
of ``odd'' functions satisfying
the relation $\tilde{f}(g) + \tilde{f}(g^{-1}h(\pi)))=0$. Therefore, restricted to the subspace $Fun(SU(2))/Ker \simeq Fun(SO(3))$ 
the map is an isomorphism and its inverse is given by \cite{FreidelLivine}:
\begin{eqnarray}
f \; \longmapsto \; \tilde{f} \; :  \; g \; \longmapsto \; \tilde{f}(g) \; = \; \frac{1}{8\pi} \int d^3\vec{x} \; 
(f \star w_{g^{-1}})(\vec{x}) \;.
\end{eqnarray}

\vspace{0.3cm}

{\it d. Symmetry transformations of the bosonic states.}

\medskip

Before going to the construction of the quantum field, we need to write the action of elements $(\vec{x},u) \in DSU(2)$ 
on bosonic states. To do so, we first compute the following coproduct:
\begin{eqnarray}
\Delta^{(n)}(\vec{x},u) \; = \; \int \prod_{i=1}^ndg_i \; w_{g_1 \cdots g_n}(\vec{x}) \; \bigotimes_{i=1}^n (g_i,u) 
\end{eqnarray}
which is an immediate consequence of the expression (\ref{DSU(2)co}).
Then, the action of elements $(\vec{x},u) \in DSU(2)$ on a (non-symmetrized) one-particle states tensor product 
$\vert \otimes_i \lambda_i;\gamma_n> \in {\cal H}_n[m;\gamma_n]$ reads:
\begin{eqnarray}\label{DSU(2)symmetry}
U(\vec{x},u) \vert \otimes_i \lambda_i;\gamma_n> & = &  
\pi_m^{\otimes n} \Delta^{(n)}(\vec{x},u) \vert \otimes_i \lambda_i;\gamma_n> \nonumber \\
 & = & w_{\lambda_1h(m)\lambda_1^{-1}} \star \cdots \star w_{\lambda_nh(m)\lambda_n^{-1}}(\vec{x}) 
\vert \otimes_i u^{-1}\lambda_i;\gamma_n> \nonumber \\
 & = & \exp[i \text{tr}(x\prod_{i=1}^n \lambda_i h(m) \lambda_i^{-1})] \vert \otimes_i u^{-1}\lambda_i;\gamma_n> \;.
\end{eqnarray}
We used the same notations as the ones introduced for the non-gravitational case. 
These transformation laws extend trivially to the 
bosonic Fock space ${\cal F}[m,\gamma]$ by symmetrization and $U(\vec{x},u)$ are still unitary operators. 
It will be useful in the sequel to distinguish
deformed translations from rotations and we will use the notations $T(\vec{x})\equiv U(\vec{x},1)$ and $R(u)=U(\vec{0},u)$.

Due to the non-commutativity nature of the spacetime $\mathbb E^3_G$, it is convenient to dualize the translations operators
and to define the translations co-actions $T^*$ on multi-particles states as follows:
\begin{eqnarray}
T^* \; : \; {\cal H}_n[m;\gamma_n]^* & \longrightarrow & {\cal H}_n[m;\gamma_n]^* \otimes Fun(\mathbb E^3_G)\\
<\lambda_1,\cdots,\lambda_n;\gamma_n \vert & \longmapsto & <T^*(\lambda_1,\cdots,\lambda_n;\gamma_n) \vert \;
\equiv \; <w_{g_1}\lambda_1,\cdots,w_{g_n}\lambda_n;\gamma_n \vert \nonumber 
\end{eqnarray}
where $g_i=\lambda_i h(m) \lambda_i^{-1}$ and
\begin{eqnarray*}
<T^*(\lambda_1,\cdots,\lambda_n);\gamma_n \vert (\vec{x}) \vert \lambda_1',\cdots,\lambda_{n'}';\gamma_{n'}>\; \equiv \;  
<\lambda_1,\cdots,\lambda_n;\gamma_n \vert T(\vec{x}) \vert \lambda_1',\cdots,\lambda_{n'}';\gamma_{n'}>
\end{eqnarray*}
We have introduced the notation ${\cal H}_n[m;\gamma_n]^*$ for the dual Hilbert space of ${\cal H}_n[m;\gamma_n]$ and used bra-ket 
notations for vectors and linear forms. The image of $T^*$ is  canonically identified with a subspace 
${\cal W}_n[m;\gamma_n] \subset {\cal H}_n[m;\gamma_n] \otimes Fun(\mathbb E^3_G)$ which naturally generalizes the space of 
$n$-particles states. The space ${\cal W}_n[m;\gamma_n]$ puts forward the phases induced by translations on $n$-particles states:
in the deformed case, usual phases are replaced by functions on the $\mathbb E^3_G$.
The definition of $T^*$ trivially extends to the space of symmetric states and we will denote the corresponding image by
${\cal W}^s_n[m;\gamma_n]$. 
By convention, we will adopt the notation $\vert w_g \lambda_1 \otimes w_0 \lambda_2> = \vert w_g \lambda_1 \otimes \lambda_2>$
for elements on ${\cal W}_n[m;\gamma_n]$ and similarly for elements of ${\cal W}^s_n[m;\gamma_n]$.

\medskip

Let us finish this section by a remark. Pure momenta states are no-longer (compared to the
non-gravitational case) eigenvectors of the translation operators $U(\vec{x}) \equiv U(\vec{x},1)$ due to the non-co-commutativity 
of $DSU(2)$. For instance, a 2-particles bosonic state transforms as:
\begin{eqnarray*}
T(\vec{x}) \vert \lambda_1 \otimes^s \lambda_2>  \; = \;  
e^{i \text{tr}(x \lambda_1 h(m) \lambda_1^{-1}\lambda_2 h(m) \lambda_2^{-1}) }\vert \lambda_1 \otimes \lambda_2> + 
e^{i \text{tr}(x \lambda_2 h(m) \lambda_2^{-1} \lambda_1 h(m) \lambda_1^{-1})}\vert \lambda_2 \otimes \lambda_1>.
\end{eqnarray*}
The resulting state is still symmetric but there is an interference term that prevents it from being a pure momentum state.

\vspace{0.3cm}

{\it e. The Quantum Field operator.}

\medskip

Bosonic creation and annihilation operators for the self-gravitating field are denoted as usual $a^\dagger(\lambda)$ and $a(\lambda)$
respectively and satisfy the same commutation relations as the non-gravitating field (\ref{commutationrelations}). They act on the 
bosonic Fock space (\ref{Fockspace}) by raising and lowering the number of particles according to the following maps:
\begin{eqnarray}\label{creann}
a^\dagger(\lambda)  & : &  {\cal H}_n^s[m;\gamma_n]  \longrightarrow  {\cal H}_{n+1}^s[m;\gamma_{n+1}] \;,\;\;
\vert \otimes_i^s \lambda_i;\gamma_n>  \longmapsto  \vert \otimes_i^s \lambda_i \otimes^s \lambda;\gamma_{n+1}> \\
a(\lambda)  & : & {\cal H}_n^s[m;\gamma_n]  \longrightarrow  {\cal H}_{n-1}^s[m;\gamma_{n-1}] \;,\;\;
\vert \otimes_i^s \lambda_i;\gamma_n> \longmapsto  \sum_{r=1}^n \delta(\lambda^{-1} \lambda_r) \vert \otimes_{i \neq r}^s\lambda_i
;\gamma_{n-1}>.
\end{eqnarray}
Note that, whatever the value of $\lambda$, $a(\lambda)$ deletes always the link
$\ell_{n}$ of $\gamma_n$ and then there is a redistribution of the momenta on the remaining links. 
It will be convenient to introduce the notations $a_+(\lambda)$ and $a_-(\lambda)$ respectively for creation and annihilation 
operators.

Symmetries on the states induce symmetries on the operators $a_\pm(\lambda)$. Rotations work in the same way as for the 
classical case and we have:
\begin{eqnarray}
R(u) \rhd a_\pm(\lambda) \; = \; a_\pm(u^{-1} \lambda) \;.
\end{eqnarray} 
This action satisfies
$R(u)(a_\pm(\lambda) \vert \otimes_i \lambda_i;\gamma_n>) = (R(u) \rhd a_\pm(\lambda)) (R(u)\vert  \otimes_i \lambda_i;\gamma_n>)$
and therefore it is customary to write $R(u) \rhd a_\pm(\lambda)=R(u) a_\pm(\lambda) R(u)^{-1}$.

Translations are a bit more involved to define on creation and annihilation operators. We start by 
denoting $\cal O$ the set of operators on the Fock space defined as a finite product of creation and annihilation operators.
The co-action $T^*$ extends to the space $\cal O$ and we have:
\begin{eqnarray}
T^* \rhd a^\dagger(\lambda) & \equiv & a^\dagger(w_{g^{-1}} \lambda) \;\;\; \text{with} \;\; g=\lambda h(m) \lambda^{-1} \;\;\; 
\text{and s.t.} \\
&& a^\dagger(w_g \lambda) \vert \otimes_i^s \lambda_i;\gamma_n> = 
\vert \otimes_i^s \lambda_i \otimes^s w_g \lambda;\gamma_n> \in {\cal W}_{n+1}[m;\gamma_{n+1}] \;.
\end{eqnarray}
The action on $a(\lambda)$ is obtained by adjointness and extends to the whole set $\cal O$ by linearity
and morphism, i.e.
$T^*\rhd (a^\dagger(\lambda) a^\dagger (\lambda_2)) = T^*\rhd (a^\dagger(\lambda)) T^*\rhd (a^\dagger (\lambda_2))$.
Then, we see immediately that we have the consistency relation:
\begin{eqnarray}
<T^*(\otimes_i^s \lambda_i);\gamma_n \vert \; = \; <T^* \rhd (\prod_{i=1}^n a^\dagger(\lambda_i))  0 \vert  \;
\end{eqnarray}
where $<0\vert$ is the vaccum state. Moreover, given an element $A \in \cal O$ and a state $<s\vert$, we have the property:
\begin{eqnarray}
<T^*(A s)\vert \; = \; \sum_{(T^*)} <(T^*_{(1)} \rhd A) (T^*_{(2)} s) \vert \; = \; <(T^* \rhd A) (T^* s) \vert
\end{eqnarray}
because $\Delta_F(T^*) = T^*\otimes T^*$. Therefore, as for the classical case, it is consistent to write formally the action 
of $T^*$ on $A$ as $T^* \rhd A = T^* A S(T^*)$ which means that:
\begin{eqnarray}\label{TTT}
(T^* \rhd A)(\vec{x}) = \sum_{(\vec{x})}(T^*(\vec{x}_{(1)})) A (S(T^*)(\vec{x}_{(2)}))\;.
\end{eqnarray}
This identity is the deformed analog of the classical relation $T(\vec{x}) \rhd A = T(\vec{x}) A T(-\vec{x})$.
In that sense, we say that creation and annihilation operators transform covariantly
under translations.

\medskip

Now, we have all the ingredients to construct the quantum field operator for a massive self-gravitating quantum field theory.
Such a field $\phi$ is defined as Fock space operator valued function on $\mathbb E^3_G$ that satisfies the properties of locality,
covariance and causality. We consider the self-adjoint state at the origin defined as
\begin{eqnarray}
\phi(\vec{0}) \; \equiv \; \int d\mu(\lambda) \; (c_+(\lambda) a^\dagger(\lambda) + c_-(\lambda)a(\lambda)) 
\end{eqnarray}
where $c_\pm$ are complex valued functions on $S^2$. If we ask that $\phi(\vec{0})$ is invariant under rotations $R(u)$,
then the functions $c_\pm$ are in fact fixed to a constant $A_\pm$ that can be chosen real. The field $\phi(\vec{x})$
at any point $\vec{x} \in \mathbb E^3_G$ is obtained by covariance, i.e. $\phi \equiv T^* \phi(\vec{0})$ and then:
\begin{eqnarray}
\phi(\vec{x}) \; = \; \int d\mu(\lambda) (A_+ e^{i\text{tr}(x\lambda h(m)\lambda^{-1})} a^\dagger(\lambda) +
A_- e^{-i\text{tr}(x\lambda h(m)\lambda^{-1})} a(\lambda)). 
\end{eqnarray}
As for the classical case, the causality requirement is meanningless in an euclidean theory. All the same, we say that the field is
causal if $[\phi(\vec{x}),\phi(\vec{y})]=0=[\phi(\vec{x}),\phi(\vec{y})^\dagger]$ which implies that $A_+=A_-=A_G$. The constant $A_G$
will be fixed later on. 

At this point, we can compute the propagator of the self-gravitating quantum field theory given by the two points function:
\begin{eqnarray}
\Delta_G(\vec{x},\vec{y}) \; = \; A_G^2 \int d\mu(\lambda) \; e^{i\text{tr}((x-y)\lambda h(m) \lambda^{-1})} \; = \; 
A^2_G \frac{\sin (\vert \! \vert \vec{x} - \vec{y} \vert \! \vert \sin m)}{\vert \! \vert \vec{x} - \vec{y} \vert \! \vert \sin m} \;.
\end{eqnarray}
Quantum gravity is responsible of the renormalization of the mass of the quantum field which is no-longer given by $m$ but by
$\sin m$. Note that the mass is expressed in term of the Planck mass $m_p=1/G$ and the distances are expressed in term of the
Planck length $l_p=G$ ($\hbar$ and $c$ are set to one). 

At the non-gravity limit $G \mapsto 0$, 
 $\Delta_G$ tends to the classical propagator $\Delta$ according to the following limit:
\begin{eqnarray} \label{QGcorrectionstopropagator}
\Delta_G(\vec{x}) \; = \; A_G^2 \frac{\sin(\frac{\vert \! \vert \vec{x} \vert \! \vert}{l_p}\sin(\frac{m}{m_p}))}{
\frac{\vert \! \vert \vec{x} \vert \! \vert}{l_p} \sin \frac{m}{m_p}} \; \simeq \; \frac{A_G^2}{A^2} \Delta_0(\vec{x}) 
\; (1-\frac{m^2 G^2}{6}\left(m\vert \! \vert \vec{x} \vert \! \vert \text{cotan} (m\vert \! \vert \vec{x} \vert \! \vert )-1)\right) 
+ {\cal O}(G^3).
\end{eqnarray}
The choice $A_G=A=1$ is consistent with the classical limit.

Note that $\Delta_G$ is in fact the symmetric propagator associated to $DSU(2)$ as defined in \cite{FNR}.
A symmetric propagator is a function on $\mathbb E^3_G = DSU(2)/SU(2)$ defined by a simple representation labelled by a mass $m$
as follows:
$$
K_m(\vec{x})= \int d\mu(\lambda) \; \overline{\omega(\lambda)} (\pi_m(\vec{x}) \omega)(\lambda) 
$$ 
where $\omega \in Fun(S^2)$ is the unique $SU(2)$ co-invariant function. A direct comparison of the previous formula and the
formula defining $\Delta_G$ shows the equality between the two functions. 
It may be interesting to show this equality starting from the QFT two-points function as follows:
\begin{eqnarray}
\Delta_G(\vec{x}) & = & <0\vert \phi(\vec{0}) \phi(\vec{x})\vert 0> \nonumber \\
& = & \sum_{(\vec{x})}<0\vert \phi(\vec{0})(T^*(\vec{x}_{(1)}))  \phi(\vec{0}) (S(T^*)(\vec{x}_{(2)})) \vert 0> \nonumber\\
& = & <\omega \vert T^*(\vec{x}) \omega > \; = \; <\omega \vert \pi_m(\vec{x}) \omega> \; = \; K_m(\vec{x})\;.\nonumber 
\end{eqnarray}
We have succesively used the relation (\ref{TTT}) for symmetry transformations of the quantum fields, the fact that 
$\omega(\lambda)=<\lambda \vert \phi(\vec{0}) 0>$ is the normalized $SU(2)$ invariant function on $DSU(2)$ and finally
the invariance of the vacuum state $\vert 0>$ under translations. Moreover, the representation $m$ appears in the last line of
previous calculations for the field is implicitely of mass $m$.

\medskip

Finally, the free self-gravitating field in not so different than the free non-gravitating scalar field theory: the main difference is
the mass term that becomes bounded for the usual mass $m$ is replaced by $\sin(m/m_p)$ where $m_p$ is the Planck mass. This result
is consistent with those obtained in different context like spin-foam \cite{FreidelLivine} or combinatorial quantization \cite{BKM}.
The theory becomes more interesting when one introduces self-interaction in the model. This is what we are going to do in the next section.

\subsection*{3. Self-interacting self-gravitating quantum field}
The previous section was devoted to the construction of a free self-gravitating quantum scalar field theory. By free, we mean that
the field is only subjected to gravitational interactions but not to self-interaction. This section aims at introducing self-interactions
in the theory and computing transition amplitudes. To do so, we leave the hamiltonian formulation for a moment and switch into the
lagrangian formulation.

\subsubsection*{3.1. The lagrangian formulation}
We want to construct the free lagrangian theory whose propagator gives back the Hadamard propagator we found in the hamiltonian
framework. 
As we shown in the previous section, quantum states of a self-gravitating spinless scalar field of mass $m$ are described in term of
functions $\phi$ on the quantum group $DSU(2)$ which are co-invariant under the action of $SU(2)$. Moreover, the propagator is defined
as the symmetric intertwiner $K_m$ and therefore it is natural to view the action
defining the lagrangian of the free field as the following integral on $DSU(2)^{\otimes 2}$:
\begin{eqnarray}
S_0[\phi] \; \equiv \; h^{\otimes 2}(\phi_1 \; K_{12}^{-1} \; \phi_2 ) \;\;\;\; 
\end{eqnarray}
where $h: DSU(2) \rightarrow \mathbb C$ is the Haar measure on $DSU(2)$ and we use the standard universal notations 
$\phi_1=\phi \otimes id$ and so on. The inverse kernel $K^{-1}$ is defined such that 
\begin{eqnarray}
(\text{Id} \otimes h \otimes \text{Id}) K_{12}^{-1} \; K_{23} \; = \; 
(\text{Id} \otimes h \otimes \text{Id}) K_{12} \; K_{23}^{-1} \; = \; \text{Id}_{13} \;. 
\end{eqnarray}
When restricted on mass $m$ particles states, $K$ is in fact the identity and therefore is its own inverse 
($K$ can be viewed as a projector \cite{Krasnov}).
One can immediately write this action as an integral on $(\mathbb E^3_G)^{\otimes 2}$:
It is also interesting to view the action as an integral on $C_m^{\otimes 2}$ where $C_m$ is the conjugacy class defined
by $C_m \equiv \{g \in SU(2) \vert \exists \; \lambda \in S^2, \; g = \lambda h(m) \lambda^{-1}\}$ and physically represents
the space of momenta of one-particle states. The normalized measure
on $C_m$ will be denoted $d\mu_m(g)$. Therefore, in term of positions or momenta variables, the free field action reads:
\begin{eqnarray}
S_0[\phi] & = & \int d^3\vec{x} \; (\phi \star \phi )(\vec{x}) \; = \; \int d^3\vec{x} d^3\vec{y} \; \phi(\vec{x}) 
\Delta_G(\vec{x}-\vec{y}) \phi(\vec{y}) \\
& = & \int d\mu_m(g) \; \widetilde{\phi}(g) \; \widetilde{\phi}(g^{-1}) 
\end{eqnarray}
where $\phi$ and $\widetilde{\phi}$ are related by the Fourier transform $\phi = \int d\mu_m(g) \; 
\widetilde{\phi}(g) \; w_g$. Thus, the action of the self-gravitating quantum field theory can be viewed 
as a non-commutative quantum field theory, a non-local quantum field theory or a group field theory. 
These three pictures are 
completely equivalent and one can see the non-commutativity (or the non-locality) as the result of quantum gravity effects.
Whatever the formulation we consider, it is immediate to show the identity:
\begin{eqnarray}
\int [{\cal D}\phi] \; \phi(\vec{x}) \; \phi(\vec{y}) \; e^{iS_0[\phi]} \; = \; <0\vert \phi(\vec{x}) \phi(\vec{y})\vert 0>  
\end{eqnarray}
which prooves the equivalence between the lagrangian and hamiltonian descriptions. As usual, $[{\cal D}\phi]$ is the normalized
path integral measure.

Note that there is no dynamics (i.e. no kinetic term) in the theory for we deal with euclidean theory without causal structure.
Indeed, from the very construction of our theory, the two-points function is in fact the deformed analog of the Haadamard propagator
which is a solution and not a Green function of the equations of motion. To recover a dynamic, one can replace the propagator
$\Delta_G$ by its associated Green function as it was done in \cite{FreidelLivine}. In that case, one obtain the following dynamical 
lagrangian (up to a global constant):
\begin{eqnarray*}
S_0[\phi] \; = \; \int d^3\vec{x} \; 
\left[ \partial_i \phi \star \partial_i \phi \; - \; \sin^2 (m) (\phi \star \phi)(\vec{x})\right]\;.
\end{eqnarray*}
Recently, it was nicely shown \cite{OritiTlas} that the emergence of the dynamical part of the action can be viewed as
a consequence of the implementation of the causality in the context of spin-foam models.
This is the most natural way to have a dynamical
theory but the quantum evolution becomes non-unitary which is a recurrent problem of non-commutative quantum field theories
\cite{FreidelLivine}. Apparently\footnote{Private communication with E. Livine.}, 
the problem of unitarity is solved when one deals with lorentzian theory instead. 
As we deal with euclidean theory, we keep studying the non-dynamical theory that reproduces in fact the non-causal spin-foam models.
This is the reason why we call our theory non-causal. 

\medskip

Up to now, we have concentrated only on the case of the free field theory. Therefore, we do not have non-trivial transition
amplitudes involving particles creation and annihilation processes in our model. In order to include such transition amplitudes,
we need to add an interaction term $S_{int}$ to the action $S_0$. The only physical requirement we ask is that $S_{int}$ is invariant 
under the action of $DSU(2)$ as $DSU(2)$ appears to be the symmetry quantum group of our field theory. Furthermore, we restrict
ourselves to a tri-valent interaction. As a consequence, the interaction term $S_{int}$ is a three valent intertwining
coefficient given by:
\begin{eqnarray}
S_{int}[\phi] \; \equiv \; \frac{\lambda}{3!} \; h(<\omega^{\otimes 3} \vert  \; \iota \; \vert \phi \otimes \phi \otimes \phi>)
\;\;\;\text{with} \;\; 
\iota(\vec{x}) \; = \; (\pi_m\otimes \pi_m\otimes \pi_m) \Delta^{(3)}(\vec{x}) \;.
\end{eqnarray}
Note that $<\omega^3 \vert \iota$ is the symmetric intertwiner between three same simple representations labelled by $m$ that takes
value in the space of functions on $\mathbb E^3_G \subset DSU(2)$; recall that $\omega$ is the normalized $SU(2)$ invariant vector 
and $h$ the Haar measure on $DSU(2)$. To be more concrete, one can write the interacting term in positions
or momenta variables as follows:
\begin{eqnarray}
S_{int}[\phi] & = & \frac{\lambda}{3!} \int d^3\vec{x} \; (\phi \star \phi \star \phi)(\vec{x}) \\
              & = & \frac{\lambda}{3!} \int d\mu_m(g_1) \; d\mu_m(g_2) \; d\mu_m(g_3) \; 
\delta(g_1g_2g_3) \; \tilde{\phi}(g_1) \;  \tilde{\phi}(g_2) \; \tilde{\phi}(g_3)
\end{eqnarray}
As for the free action $S_0[\phi]$, the interacting term can be written as a non-local interaction as shown by the following
expression:
\begin{eqnarray}
S_{int}[\phi] & = & \int d^3\vec{x} \; d^3\vec{y} \; d^3\vec{z} \; V_{int}(\vec{x},\vec{y},\vec{z}) \; {\phi}(\vec{x}) \;  
{\phi}(\vec{y}) \; {\phi}(\vec{z}) \\
& \text{with} & V_{int}(\vec{x},\vec{y},\vec{z}) \; = \; \frac{\lambda}{3!}
\int d\mu_m(g_1) \; d\mu_m(g_2) \; w_{g_1}(\vec{x}) \; w_{g_2}(\vec{y}) \; w_{g_1g_2}(-\vec{z}) \;.
\end{eqnarray} 
At the non-gravity limit ($G \rightarrow 0$), it is clear that the interaction becomes commutative. 

\medskip

The theory defined by the total action $S[\phi] = S_0[\phi] + S_{int}[\phi]$ has been introduced and studied in the context
of spin-foam models in \cite{FreidelLivine}.
From our point of view, $S[\phi]$ is clearly invariant under the action of $DSU(2)$ because it is contructed from $DSU(2)$
intertwiners: in fact, the quadratic term $S_0$ is a two-valent intertwiner and the cubic term $S_{int}$ is obviously, by construction, 
a three-valent intertwiner. 
In order to show the invariance explicitely, we compute $U(\vec{x},u) S[\phi]$, i.e. how the action $S[\phi]$ transforms under 
the action of an element $(\vec{x},u) \in DSU(2)$. For that purpose, it is more convenient to work in the momenta representation 
(the group field version of the action) and we have:
\begin{eqnarray}
U(\vec{x},u)\; S_0[\phi] & = & \int d\mu_m(g) \; 
(\pi_m^{\otimes 2} \Delta(\vec{x},u) \; \tilde{\phi} \otimes \tilde{\phi})(g\otimes g^{-1}) 
\; = \; \epsilon(\vec{x}) \; S_0[\phi] \nonumber \\
U(\vec{x},u)\; S_{int}[\phi] & = & \int d\mu_m(g_1)d\mu_m(g_2)d\mu_m(g_3) \; \delta(g_1g_2g_3)
(\pi_m^{\otimes 3} \Delta^{(3)}(\vec{x},u)  \; \tilde{\phi} \otimes \tilde{\phi} \otimes \tilde{\phi})(g_1 \otimes g_2 \otimes g_3) 
\nonumber \\
& = & \epsilon(\vec{x}) \; S_{int}[\phi].\nonumber 
\end{eqnarray}
where $\epsilon(\vec{x})=1$ is the co-unit on $\mathbb E^3_G$. The action transforms under the trivial representation of
$DSU(2)$ and is, as a consequence, $DSU(2)$-invariant. The action $S[\phi]$ is also invariant under braidings. Indeed, the braiding
action on the space of multi-particles states induces a braiding map at the level of the action defined by the following:
\begin{eqnarray}
S_0[\phi]     & \equiv & \tilde{S}_0[\phi \otimes \phi] \; \longmapsto \; \tilde{S}_0[R(\phi \otimes \phi)] \nonumber \\
S_{int}[\phi] & \equiv & \tilde{S}_{int}[\phi \otimes \phi \otimes \phi] \; 
\longmapsto \; \tilde{S}_{int}[R(\phi \otimes \phi) \otimes \phi]\nonumber
\end{eqnarray}
where we have emphasized the fact that $S_0$ is quadratic whereas $S_{int}$ is cubic in the field and $R$ denotes the action
of the $DSU(2)$ $R$-matrix. This invariance is in fact a direct consequence of the fact that symmetric intertwining coefficients are left
invariant under braidings (see \cite{FNR} for example).
Note that there is an apparent ambiguity in the definition of the braiding action on $S_{int}$ for one can act on any pair of the three
arguments defining $S_{int}$. In fact, one can easily show that the braiding action is unchanged whatever the choice of the pair of
arguments.
A corollary of these invariances is that transitions amplitudes are invariant under $DSU(2)$ and under
braiding as expected from the begining.

\medskip

Before computing examples of transition amplitudes, we generalize our model to the case where the mass $m$ of the field is not
fixed. Such a generalization is immediate for one only has to relax the condition that the field $\phi$ is a function
on the conjugacy class $C_m$ and to allow the field to be a function on the whole group $SU(2)$. The ``dynamics'' of this theory 
is governed by the following group field action (in momenta variables):
\begin{eqnarray}
S_T[\tilde{\phi}] \; = \; \int dg_1 \; dg_2 \; \tilde{\phi}(g_1) \; \delta(g_1g_2) \; \tilde{\phi}(g_2) \; + \; 
\frac{\lambda}{3!} \int dg_1\;dg_2\;dg_3 \; \delta(g_1 g_2 g_3) \; \tilde{\phi}(g_1) \; \tilde{\phi}(g_2) \; \tilde{\phi}(g_3) 
\end{eqnarray}
where $dg$ is the $SU(2)$ Haar measure.
In fact, this action describes a coupling between scalar fields of different masses. It is therefore possible to compute from it 
creations and annihilations amplitude transitions involving particles of different masses. This action is exactly the one found in
the context of spin-foam models in \cite{FreidelLivine}.

\subsubsection*{3.2. Example of transition amplitudes}
This section is devoted to study some properties of our self-gravitating self-interacting 
quantum field theory threw some concrete examples.
In particular, we will focus on the computation of the propagator (two-points functions) and of the vertex (three points functions) 
first order terms (in the coupling $\lambda$). One could see quantum gravity effects on the field propagator and on some
physical processes involving particles creations and annihilatons.

\begin{figure}[h]
\psfrag{=v}{$=\;\; \delta(g_1 g_2 g_3) \;\;\; \text{or} \;\;\; N\delta^{(3)}(\vec{p}_1+\vec{p}_2 + \vec{p}_3)$}
\psfrag{=p}{$= \;\; \delta(g_1^{-1} g') \delta_m(g)\;\;\; \text{or} \;\;\; \frac{\delta(p-m)}{P(m)} \delta^{(3)}(\vec{p} -\vec{p}')$}
\psfrag{g1}{$g_1$}
\psfrag{g}{$g$}
\psfrag{gp}{$g'$}
\psfrag{g2}{$g_2$}
\psfrag{g3}{$g_3$}
\psfrag{m1}{$m_1$}
\psfrag{m2}{$m_2$}
\psfrag{m3}{$m_3$}
\psfrag{m}{$m$}
\psfrag{x}{}
\centering
\includegraphics[scale=0.8]{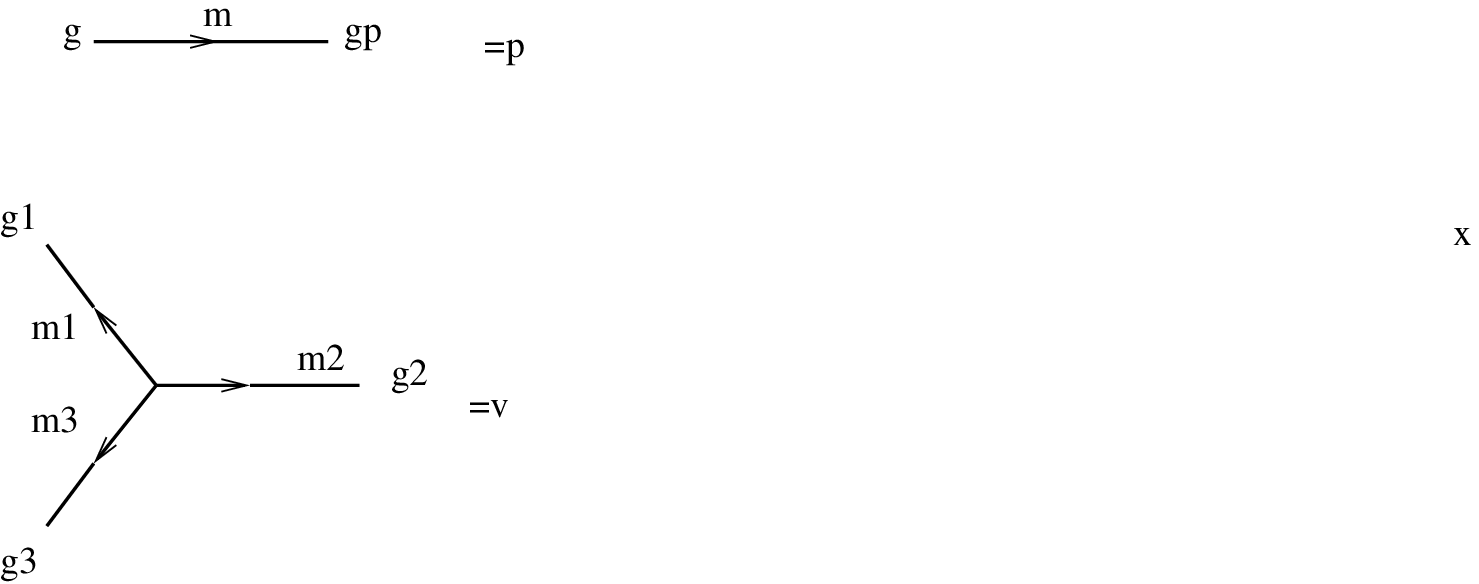}
\caption{Feynmann rules for the gravitational and classical quantum field theory (replace in that case group elements $g$ by 
vectors $\vec{p}$). 
Note that the cyclic order of the vertex is relevant for the gravitational theory but no longer for the classical one.
To be general, we have assumed that the masses of the propagators can be different.
Reversing the direction of the propagator is equivalent to change group elements by their inverse.
The propagator and vertex graphes are respectively denoted $P$ and $V$.}
\label{Feynmannrules}
\end{figure}

Before going to the details, we start by giving the Feynmann rules of the free theory in the picture of the 
(figure \ref{Feynmannrules}).
It will be convenient to compare the results of Feynmann graphs evaluations with the classical case. For that purpose, we give
the Feynmann rules for the non-gravitating quantum field theory that can be obtained by the non-gravitational limit
of the previous rules (figure \ref{Feynmannrules}). 
In particular, the normalization factors for the vertex and for the propagator are chosen such that classical
Feynmann graph evaluations fit correctly with the no-gravity limit of gravitational Feymann graph evaluations: $N=2\pi^2$ and
$P(m)=4\pi m^2$.

Given a graph $\Gamma$ we will note $I_0(\Gamma)$ and $I_G(\Gamma)$ the classical and gravitational (deformed) 
evaluations. Note that there is an overall factor $N^{-1}$ in the gravitational evaluation of any graph $\Gamma$ in the sense that
$I_G(\Gamma)=N^{-1} \tilde{I}_G(\Gamma)$ where $\tilde{I}_G(\Gamma)$ is the very Feynmann graph evaluation (according to the gravitational
Feynmann rules).

\vspace{0.3cm}

{\it a. Corrections to the propagator}

\medskip

Quantum gravity corrections to the free propagator have already been computed in the previous section (\ref{QGcorrectionstopropagator}).
This section aims at illustrating quantum gravity effects on the propagator of the self-interacting theory. Formally, the propagator
of the self-interacting theory is given by a power series in the coupling constant $\lambda$ as follows:
\begin{eqnarray}\label{formal2points}
<\phi(\vec{x}_1) \; \phi(\vec{x}_2)> \; \equiv \; 
\sum_{\Gamma} \lambda^{v(\Gamma)} \frac{1}{\text{Sym}(\Gamma)} F_{(\Gamma)}(\vec{x}_1,\vec{x}_2)\;.
\end{eqnarray}
The sum runs over all Feynmann graphs $\Gamma$ with two open edges, $v(\Gamma)$ is the number of vertices, $\text{Sym}(\Gamma)$ 
the symmetry factor and $F_{(\Gamma)}$ is the evaluation of the graph viewed as a function of the position variables $\vec{x}_1$ 
and $\vec{x}_2$.
Here we are not interested in convergence and renormalization issues and we concentrate only on lowest order terms in the series.  
The lowest term $F_P$ ($P$ being the trivial graph that contains only one edge) 
is the free propagator and is given by the function $\Delta_I(\vec{x}_1,\vec{x}_2)$ computed in previous
sections ($I$ is $0$ or $G$ if we are dealing with the non-gravitating or the self-gravitating theory).

\begin{figure}[h]
\psfrag{m}{$m$}
\psfrag{g1}{$g_1$}
\psfrag{g2}{$g_2$}
\psfrag{g3}{$g_3$}
\psfrag{m1}{$m_1$}
\psfrag{m2}{$m_2$}
\psfrag{m3}{$m_3$}
\psfrag{m4}{$m_4$}
\psfrag{m5}{$m_5$}
\psfrag{m6}{$m_6$}
\centering
\includegraphics[scale=0.8]{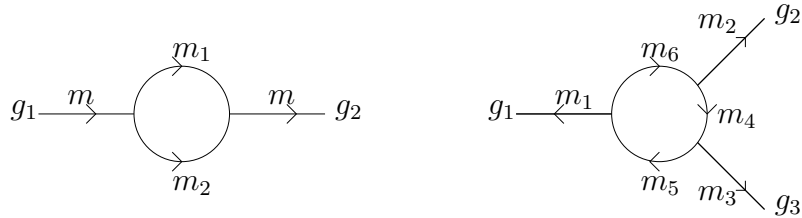}
\caption{One loop corrections of the propagator and of the vertex: 
the graphes are respectively denoted $\Lambda$ and $T$. 
In-going and out-going particles are pure momenta states
whose momenta are the group elements $g_1$, $g_2$ and $g_3$.}
\label{oneloop}
\end{figure}

Quantum gravity corrections to higher order terms in the series (\ref{formal2points}) are a bit more involved to compute. We want
to illustrate them with the non-trivial example of the Feynmann graph $\Lambda$ drawn in the figure (\ref{oneloop}).

To be general, we assume that in-going and out-going particles have the same mass $m$ whereas the particles inside the loop have
different masses fixed to the values $m_1$ and $m_2$. 
An immediate calculation shows that 
$F_{(\Lambda)} \equiv f^I_{(\Lambda)}(m,m_1,m_2) \Delta_I(\vec{x}_1,\vec{x}_2)$. In the gravitational case, the coefficient
$f^G_{(\Lambda)}$ reads:
\begin{eqnarray}
f^G_{(\Lambda)}(m,m_1,m_2) & \equiv & \int d\mu_{m_1}(g_1) \; d\mu_{m_2}(g_2) \; \delta(g^{-1}g_1 g_2) \\
& = & \sum_{\ell =0}^{\infty} \frac{1}{\ell} \frac{\sin(\ell m) \; \sin( \ell m_1) \; \sin(\ell m_2)}{\sin(m) \; \sin(m_1) \; \sin(m_2)} \;
=\; \frac{\pi}{4}\frac{Y(m_1,m_2,m_3)}{\sin m_1 \; \sin m_2 \; \sin m_3}. \nonumber
\end{eqnarray}
$Y(m_1,m_2,m_3)$ is equal to one if $m_1,m_2,m_3$ satisfy triangular inequalities and is equal to zero otherwise.
It is interesting to compare this expression to the classical expression $f^0_{(\Lambda)}$.
A similar calculation shows that:
\begin{eqnarray}
f^0_{(\Lambda)}(m,m_1,m_2) & \equiv & N^2 \int \frac{d^3\vec{p}_1}{P(m_1)} \; \frac{d^3\vec{p}_2}{P(m_2)} \; 
\delta^{(3)}(\vec{p}_1+\vec{p}_2 +\vec{p}) \\
 & = &  \frac{N^2}{2\pi^2} \int \frac{dx}{x} \frac{\sin (xm) \; \sin(xm_1) \; \sin (xm_2)}{m_1 \; m_2 \; m_3} \; = \; 
\frac{\pi}{4}\frac{Y(m_1,m_2,m_3)}{m_1 \; m_2 \; m_3}. \nonumber
\end{eqnarray}
We have denoted by $\vec{p}$ the in-going (and out-going) momentum whose mass is $m$. 
Note that the expressions of $f^G_{(\Lambda)}$ and $f^0_{(\Lambda)}$ are very similar: 
the first is expressed as a discrete series whereas the second is
an integral over the space variable $x$. This suggests, as expected, that space-time becomes discrete at the Planck scale (at least
for euclidean gravity) and the spectrum of the length variable $x$ is given by $(2I+1)l_p$. 

For the massive quantum field, we are exclusively interested to the case where all the masses are fixed to the same value
$m$. In that case, the first order gravitational corrections to the Feynmann graph $\Lambda$ are trivially obtained
from quantum gravity corrections of the free propagator and from the following corrections:
\begin{eqnarray}
f^G_{(\Lambda)}(m) \; \equiv \; f^G_{(\Lambda)}(m,m,m)  =  
\frac{\pi}{4 \sin^3m} \; = \; \frac{1}{G^3} \; (1 + \frac{m^2G^2}{2} + {\cal O}(G^3)) \; f^0_{(\Lambda)}(m,m,m) \; . 
\end{eqnarray}
Note the presence of the overall factor $G^{-3}$ in the expansion of $f^G_{\Lambda}(m)$; then the coefficient that admits
the right classical limit is $G^{3}f^G_{\Lambda}(m)$. 
In the hamiltonian point of view, this means that the two-points function for the interacting quantum field theory is modified
du to the quantum gravity effects and is given (at the lowest order in the couplings $\lambda$ and $G$) by:
\begin{eqnarray}
<0\vert \phi(\vec{0}) \; H_{int} \; \phi(\vec{x})> \; = \; (1+ G^{3}\frac{\lambda^2}{2} f^G_{(\Lambda)}(m) 
+{\cal O}(\lambda^3)) \Delta_G(\vec{x})  \;.
\end{eqnarray}
The formal notation $H_{int}$ is for the self-interaction of the self-gravitating quantum field. Quantum gravity effects 
appears in the expression of the free propagator $\Delta_G(\vec{x})$ and also in the expression of $\Delta_G^{(1)}$, i.e. the perturbative
corrections du to self-interaction. 

\vspace{0.3cm}

{\it b. Corrections to the three-points function}

\medskip

The three-points function $<\phi(\vec{x}_1)\phi(\vec{x}_2) \phi(\vec{x}_3)>$ is defined as a series in the coupling 
constant $\lambda$  which can be formally writen as follows:
\begin{eqnarray} \label{formal3points}
<\phi(\vec{x}_1)\phi(\vec{x}_2)\phi(\vec{x}_3)> \; \equiv \; \sum_{\Gamma} \lambda^{v(\Gamma)}\frac{1}{\text{Sym}(\Gamma)} 
H_{(\Gamma)} (\vec{x}_1,\vec{x}_2,\vec{x}_3)
\end{eqnarray}
where the sum runs over Feynmann graphs $\Gamma$ with three open edges. 
As for the previous section, we are particularly interested in computing some lowest order terms $H_{(\Gamma)}$ 
in the series: the evaluation of the graph $V$ is the vertex of the free theory in the position variables, the evaluation
of the tetrahedron graph $T$ is a lowest order correction to the self-interacting theory vertex.
We will introduce the index $I \in \{0,G\}$ in the functions $H^I_{(\Gamma)}$ of the series (\ref{formal3points})
to specify the ``classical'' and the self-interacting theories.

\medskip

Let us start by computing $H_{(V)}^0(\vec{x}_1,\vec{x}_2,\vec{x}_3)$. A direct application of Feynmann rules leads to the 
following expression:
\begin{eqnarray}\label{tree3points}
H^0_{(V)}(\vec{x}_1,\vec{x}_2,\vec{x}_3) \; \equiv \; \int \prod_{\ell=1}^3 d^3\vec{p}_\ell\; 
\frac{\delta(p_\ell-m_\ell)}{P(m_\ell)} \; e^{i\vec{p}_\ell \cdot \vec{x}_\ell} \; \delta^{(3)}(\vec{p}_1+\vec{p}_2+\vec{p}_3)
\end{eqnarray}
Note that we keep different masses of out-going and in-going particles to be more general. Therefore, the function $H^0_{(V)}$ depends also
on the parameters $m_1,m_2$ and $m_3$. It is clear from the previous expression that $H^0_{(V)}$ vanishes if $m_1,m_2$ and $m_3$ does not 
satisfy triangular inequalities and we will assume in the sequel that the masses do satisfy triangular inequalities.
In order to interpret $H^0_{(V)}$
as a tree order three points function of our quantum field theory, we have to fix the masses at the same value $m$.
In that case, it is immediate to see that $H^0_{(V)}$ satisfies the following symmetry relation:
\begin{eqnarray}\label{symprop}
H^0_{(V)}(\vec{x}_1,\vec{x}_2,\vec{x}_3)=H^0_{(V)}(\vec{x}_1-\vec{x}_3,\vec{x}_2-\vec{x}_3,\vec{0})\;.
\end{eqnarray}
This property is easy to see from  the expression (\ref{tree3points}) but it is more transparent in the following expression
that we obtain by decomposing the delta function $\delta^{(3)}(\vec{p})$ in Fourier modes:
\begin{eqnarray}
H^0_{(V)}(\vec{x}_1,\vec{x}_2,\vec{x}_3) \; = \; \frac{1}{(2\pi)^3} \frac{1}{m_1\;m_2\;m_3} \int d^3\vec{x}_0 
\frac{\sin(m_1x_{10}) \sin(m_2x_{20}) \sin(m_3x_{30})}{x_{10} \; x_{20} \; x_{30}}\;.
\end{eqnarray}
We have introduced the notation $x_{\ell 0} = \vert \!\vert \vec{x}_\ell - \vec{x}_0 \vert \! \vert$.  This expression is interesting for
it explicitely exhibits the symmetry between the position variables $\vec{x}_\ell$ but it is an indefinite three dimensional integral.
There exist other equivalent expressions that apparently break this symmetry but may be more useful because they are expressed as a 
single definite integral. To obtain such an expression, one starts by integrating over the momentum variable $\vec{p}_3$ in the expression
(\ref{tree3points}) and after some simple calculations, one shows that:
\begin{eqnarray}\label{class3}
H^0_{(V)}(\vec{x}_1,\vec{x}_2,\vec{x}_3) \; = \; {\cal N}_0 \; \int d^2\vec{n}_1 \; d^2\vec{n}_2 \; 
\delta(\vec{n}_1 \cdot \vec{n}_2 - \cos S_0) \; e^{i(\vec{A}_1 \cdot \vec{n}_1 + \vec{A}_2 \cdot \vec{n}_2)}
\end{eqnarray}
where $\vec{A}_i\equiv m_i(\vec{x}_i - \vec{x}_3)$ and $\cos S_0 \equiv (m_3^2 - m_1^2 + m_3^2)/(2m_1 m_2)$ with $S_0>0$
and the normalization factor is given by ${\cal N}_0^{-1}=4\pi m_1 m_2 m_3$.
As shown in the appendix C, the function $H^0_{(V)}$ can be simplified to the following expression:
\begin{eqnarray}\label{expressJJ}
H^0_{(V)}(\vec{x}_1,\vec{x}_2,\vec{x}_3)  =  \frac{{\cal N}_0}{4} \; \int_0^\pi d\theta \; \sin \theta \;
e^{iA_1 \cos U \cos \theta} J_0(A_1 \sin U \sin \theta) \;  e^{iA_2 \cos S \cos \theta} J_0(A_2 \sin S \sin \theta)   
\end{eqnarray}
with $A_i = \vert \! \vert \vec{A}_i \vert \! \vert$, the angle $U>0$ is defined by the relation 
$\vec{A}_1 \cdot \vec{A}_2 = A_1A_2 \cos U$ and $J_0$ is the Bessel function of the first kind (see appendix C for details). 
This expression will appear more convenient
when we compare to the self-gravitating analog. In some particular cases, one can perform the previous integral explicitely:
\begin{enumerate}
\item If $S_0=0 \Longleftrightarrow m_3^2=(m_1+m_2)^2 \Longleftrightarrow m_3=m_1+m_2$ ($m_3>0$) then:
\begin{eqnarray}\label{Sim1}
H^0_{(V)}(\vec{x}_1,\vec{x}_2,\vec{x}_3) \; = \; \frac{1}{8\pi \; m_1m_2m_3} 
\frac{\sin (\vert \! \vert m_1 \vec{x}_1 + m_2 \vec{x}_2 -m_3\vec{x}_3 \vert \! \vert)}
{\vert \! \vert m_1 \vec{x}_1 + m_2 \vec{x}_2 -m_3\vec{x}_3 \vert \! \vert}\;.
\end{eqnarray}
Note however that this case is not of a particular physical interest for we will be interested to the cases where 
the masses are all fixed to the same value (which is obviously incompatible with the condition $S_0=0$). 
\item If $U=0 \Longleftrightarrow$ $\vec{A}_1$ and $\vec{A}_2$ are colinear then:
\begin{eqnarray}\label{Sim2}
H^0_{(V)}(\vec{x}_1,\vec{x}_2,\vec{x}_3) \; = \; \frac{1}{8\pi \; m_1m_2m_3}
\frac{\sin \sqrt{\sum_{i=1}^3 m_i^2 x_i^2 + \sum_{i<j}^{k\neq i,j} (m_k^2 - m_i^2 -m_j^2)\vec{x}_i \cdot \vec{x}_j }}
{\sqrt{\sum_{i=1}^3 m_i^2 x_i^2 + \sum_{i<j}^{k\neq i,j} (m_k^2 - m_i^2 -m_j^2)\vec{x}_i \cdot \vec{x}_j }}
\end{eqnarray}
In the case where $m_1=m_2=m_3=m$ and $\vec{A}_1=\vec{A}_2$ (i.e. $\vec{x}_1 = \vec{x}_2 = \vec{x}$), 
the previous formula simplifies a bit more and reads:
\begin{eqnarray}
H^0_{(V)}(\vec{x},\vec{x},\vec{x}_3) \; = \; \frac{1}{8\pi \; m^4} \frac{\sin (m \vert \! \vert \vec{x} - \vec{x}_3\vert \! \vert)}
{\vert \! \vert \vec{x} - \vec{x}_3\vert \! \vert} \;.
\end{eqnarray} 
In particular, we see that $H^0_{(V)}(\vec{x},\vec{x},\vec{x})=H^0_{(V)}(\vec{0},\vec{0},\vec{0})$ as expected.
\end{enumerate}
Before studying the gravitational case, let us mention that one can re-express the function
$H^0_{(V)}$ as a series involving Gegenbauer polynomials and Bessel functions of half-integer order 
as shown in the appendix C.

\medskip

Let us now compute the function $H^G_{(V)}(\vec{x}_1,\vec{x}_2,\vec{x}_3)$. By definition, it is given by:
\begin{eqnarray}
H^G_{(V)}(\vec{x}_1,\vec{x}_2,\vec{x}_3) \; \equiv \; \frac{1}{N} \int \prod_{\ell=1}^3 dg_\ell \; \delta_{m_\ell}(g_\ell) 
\; e^{i\text{tr}(g_\ell x_\ell)} \; \delta(g_1g_2g_3)\;. 
\end{eqnarray}
We recall that $N=2\pi^2$. In order to make a clear comparison between $H^G_{(V)}$ and $H^0_{(V)}$, 
it is convenient to decompose the group elements
$g_\ell = \cos m_\ell \mathbb{I} + \sin m_\ell \; \vec{n}_\ell \cdot \vec{\sigma}$ where $\mathbb{I}$ and $\vec{\sigma}$ are 
respectively the identity and Pauli matrices; $\vec{n}_\ell$ is an unit vector. 
With this parametrization, the measure reduces to the following form
\begin{eqnarray} 
\delta_{m_i}(g_i) dg_i = d^2\vec{n}_i \;\;\;\;\;
\text{with} \;\;\; \delta_m(g)=\pi/\sin m \delta(\text{tr}(g) - 2 \cos m) \;. 
\end{eqnarray}
Using these results, one can show after some calculations that 
$H^G_{(V)}$ can be written in a form similar to the classical expression (\ref{tree3points}) and we have the following formula:
\begin{eqnarray}\label{expression1}
\frac{H^G_{(V)}(\vec{x}_1,\vec{x}_2,\vec{x}_3)}{\cos m_3} = \int \prod_{\ell=1}^3 d^3\vec{p}_\ell 
\frac{\delta(p_\ell - \sin m_\ell)}{P(\sin m_\ell)}e^{i\vec{p}_\ell \cdot \vec{x}_\ell} 
\delta^{(3)}(\vec{p}_3 + \vec{p}_1\cos m_2 + \vec{p}_2 \cos m_1 + \vec{p}_1 \wedge \vec{p}_2)\;.
\end{eqnarray}
Details can be found in the appendix C. It is clear that $H^G_{(V)}$ tends to $H^0_{(V)}$ at the no-gravity limit $G \rightarrow 0$
(up to some powers of $G$ we will precise later on). 
Another  
immediate constatation is that $H^G_{(V)}$ does not satisfy the property (\ref{symprop}) anymore: this can be interpreted as a consequence
of the non-commutativity of the space coordinates. In order to make a quantitative comparison with the classical counterpart $H^0_{(V)}$,
it is convenient to write $H^G_{(V)}$ is a form similar to (\ref{class3}) as follows:
\begin{eqnarray}\label{expression2}
H^G_{(V)}(\vec{x}_1,\vec{x}_2,\vec{x}_3) \; = \; 
{\cal N}_G \int d^2\vec{n}_1 \; d^2\vec{n}_1 \; \delta(\vec{n}_1 \cdot \vec{n}_2 - \cos S_G)
\; e^{i (\vec{B}_1 \cdot \vec{n}_1 + \vec{B}_2 \cdot \vec{n}_2 + \vec{C}\cdot \vec{n}_1 \wedge \vec{n}_2)}
\end{eqnarray}
where we have introduced the notations:
\begin{eqnarray*}
\vec{B}_1 \equiv \sin m_1(\vec{x}_1 - \cos m_2 \vec{x}_3) \;,\;\; \vec{B}_2 \equiv \sin m_2(\vec{x}_2 - \cos m_1 \vec{x}_3) \;,\;\;
\vec{C} \equiv -\sin m_1 \sin m_2 \vec{x}_3\;.
\end{eqnarray*}
The normalization factor is given by ${\cal N}_G^{-1}=4\pi \sin m_1 \sin m_2 \sin m_3$ and the angle $S_G>0$ is fixed
by the relation $\cos S_G = (\cos m_1 \cos m_2 - \cos m_3)/(\sin m_1 \sin m_2)$. One can make the integration over the variable
$\vec{n}_2$ and one shows that $H^G_{(V)}$ reduces to the following form:
\begin{eqnarray}\label{VGJ}
H^G_{(V)}(\vec{x}_1,\vec{x}_2,\vec{x}_3) \; = 
\; \frac{{\cal N}_G}{2} \int d^2\vec{n} \; e^{i(\vec{B}_1 + \cos S_G \vec{B}_2)\cdot \vec{n}}
J_0(\sin S_G R(\vec{n}))
\end{eqnarray}
where $R(\vec{n})=\sqrt{B^2_2 + C^2 -(\vec{B}_2\cdot \vec{n})^2 - 
(\vec{C}\cdot \vec{n})^2 + 2 \vec{B}_2\wedge \vec{C} \cdot \vec{n}}$. It is clear that (\ref{VGJ}) simplifies, when one takes 
the no-gravitational limit, and gives back the expression (\ref{expressJJ}) of $H^0_{(V)}$. 

In order to compute the classical limit and the quantum gravity corrections of $H^G_{(V)}$, we start by recalling that the masses and the
positions are  given in term of the Planck mass and the Planck length: therefore, we replace in the expression of $H^G_{(V)}$
the dimensionless variables $\vec{x}_i$ and the masses $m_i$ respectively by $\vec{x}_i/l_p$ and $m_i/m_p$ where $m_p=G^{-1}$ and 
$l_p=G$. In that case, $\vec{x}_i$ and $m_i$ become dimensionful variables.
After some calculations (whose details are presented in the appendix C), one shows that the development of $H^G_{(V)}$ 
around $G=0$ is given by:
\begin{eqnarray}
G^3 H^G_{(V)}(\vec{x}_1,\vec{x}_2,\vec{x}_3) \; = \; H^0_{(V)} + 
G \; {\cal N}_0 \; m_1m_2^2 \; H^{G(1)}_{(V)}(\vec{x}_1,\vec{x}_2,\vec{x}_3) + {\cal O}(G^2)
\end{eqnarray}
where the first order correcting term $H^{G(1)}_{(V)}$ reads:
\begin{eqnarray}
H^{G(1)}_{(V)}(\vec{x}_1,\vec{x}_2,\vec{x}_3) \; \equiv \; \int d^2\vec{n} \; e^{i(\vec{A}_1 + \cos S_0 \vec{A}_2)\cdot \vec{n}}
J_1(\sin S_0 \sqrt{A_2^2 - (\vec{A}_2\cdot \vec{n})^2}) \;
\frac{\vec{x}_2\wedge \vec{x}_3 \cdot \vec{n}}{\sqrt{A_2^2 - (\vec{A}_2\cdot \vec{n})^2}}\;. 
\end{eqnarray}
The variables have been introduced previously and $J_1$ is the first order Bessel function. 
Note that one has to rescale $H^G_{(V)}$ with a factor $G^3$ (as in the propagator case) 
in order to have a good classical limit. In that case, we remark that the 
lowest correcting term is generically proportional to $G$ whereas the lowest correction to the propagator is proportional to $G^2$.
In order to simplify the previous expression,
we choose a tri-dimensional basis where $\vec{A}_2=A_2(1,0,0)$ and, if we still denote by $U$ the positive angle between $\vec{A}_1$
and $\vec{A}_2$, we obtain that:
\begin{eqnarray}\label{VGgrad}
H^{G(1)}_{(V)}(\vec{x}_1,\vec{x}_2,\vec{x}_3) \; = \; \frac{1}{A_2}  \vec{x}_2 \wedge \vec{x}_3 \cdot \vec{\nabla}_{\vec{A}_1} I 
\end{eqnarray}
where $\vec{\nabla}_{\vec{X}}$ is the gradient with respect to the coordinates $\vec{X}$ and $I$ is an integral viewed as a function of
the vector $\vec{A}_1$. After some calculations (see appendix C for details), we show that $I$ is given by:
\begin{eqnarray}\label{firstorderQGcor}
I \; \equiv \; \int_0^1 \frac{dx}{\sqrt{1-x^2}} e^{i(A_1 \cos T + A_2 \cos S_0)x} J_0(A_1 \sin T \sqrt{1-x^2})
J_1(A_2 \sin S_0 \sqrt{1-x^2}) \;.
\end{eqnarray}
This is the more general expression of the lowest order correcting term due to quantum gravity to the classical three-points function.
One immediately remarks that the term $H^{G(1)}_{(V)}$ vanishes in some particular cases: 
if $\vec{x}_2$ and $\vec{x}_3$ are colinear or null
and also if $\vec{A}_1$ and $\vec{A}_2$ are colinear (one can see this property from the expression (\ref{VGgrad})). 
In that cases, one has to go further in the computation of the correcting terms and it is easy to see that the corrections are of order
$G^2$. We wont give the general expression of correcting term of order $G^2$. We will instead give the expression of $H^G_{(V)}$
in some simple cases from which it is easier to compute quantum gravity corrections.
\begin{enumerate}
\item If $S_G =0 \Longleftrightarrow  \cos(m_1+m_2)=\cos m_3 \Longleftrightarrow m_3=m_1+m_2 (m_3>0)$ then:
\begin{eqnarray}
H^G_{(V)}(\vec{x}_1,\vec{x}_2,\vec{x}_3) \; = \; \frac{1}{8\pi \sin m_1 \sin m_2 \sin m_3} 
\frac{\sin \vert \! \vert \sin m_1 \vec{x}_1 + \sin m_2 \vec{x}_2 - \sin m_3 \vec{x}_3\vert \! \vert}
{ \vert \! \vert \sin m_1 \vec{x}_1 + \sin m_2 \vec{x}_2 - \sin m_3 \vec{x}_3\vert \! \vert}\;.
\end{eqnarray}
This case has no  physical interest.
\item If the coupling vanishes i.e. $\vec{C}=\vec{0} \Longleftrightarrow \vec{x}_3 = \vec{0}$ (as $m_1 \neq 0 \neq m_2$) then
$H^G_{(V)}$ takes exactly the same form as $H^0_{(V)}$:
\begin{eqnarray}
H^G_{(V)}(\vec{x}_1,\vec{x}_2,\vec{x}_3) & = & \frac{{\cal N}_G}{4} \; \int_0^\pi d\theta \; \sin \theta \;
e^{iB_1 \cos U \cos \theta} J_0(B_1 \sin U \sin \theta) \nonumber \\
&& \;\;\;\;\;\;\;\;\;\;\;\;\;\;  \;\;\;\;\;\;\;\;\;\;\;\;  e^{iB_2 \cos S_G \cos \theta} J_0(B_2 \sin S_G \sin \theta)   
\end{eqnarray}
where $U>0$ is the angle between $\vec{B}_1$ and $\vec{B}_2$. Therefore, this integral can be performed explicitely
in the particular case where $\vec{B}_1$ and $\vec{B}_2$ are colinear and one obtains for $H^G_{(V)}$ the following expression:
\begin{eqnarray}
\frac{{\cal N}_G}{2} \frac{\sin \sqrt{\sum_{i=1}^2 x_i^2 \sin^2 m_i + \alpha x_3^2
+ 2\sum_{i<j}^{k \neq i,j} \vec{x}_i \cdot \vec{x}_j(\cos m_i \cos m_j - \cos m_k)}}
{\sqrt{\sum_{i=1}^2 x_i^2 \sin^2 m_i + \alpha x_3^2
+ 2\sum_{i<j}^{k \neq i,j} \vec{x}_i \cdot \vec{x}_j(\cos m_i \cos m_j - \cos m_k)}}
\end{eqnarray}
where $\alpha=\cos^2m_1+\cos^2m_2-2\cos m_1\cos m_2 \cos m_3$. This expression simplifies in the particular case where 
the masses are fixed to the same value $m$. It is important to remark that $H^G_{(V)}(\vec{x},\vec{x},\vec{x})$ does depend on $\vec{x}$
contrary to the classical case and explicitely reads:
\begin{eqnarray}
H^G_{(V)}(\vec{x},\vec{x},\vec{x}) & = & \frac{1}{8\pi \sin^3m} \frac{\sin(\beta(m) x)}{\beta(m) x} \\
\text{with}&\beta(m) &\equiv \sqrt{2-2\cos m(\cos^2 m + 3\cos m -3)}\;.\nonumber
\end{eqnarray}
This fact can be interpreted as a consequence of the non-commutativity of the space at the Planck length. 
\end{enumerate}

\medskip

We finish this section by computing the vertex one-loop correction contribution to the vertex associated to the tetrahedron graph $T$
represented in the figure (\ref{oneloop}). An immediate calculation shows that $H_{(T)} \equiv h^I_{(T)}(m_1,\cdots,m_6) 
H_{(V)}^I(\vec{x}_1,\vec{x}_2,\vec{x}_3)$: $m_1,m_2,m_3$ are the masses of out-going particles
and $m_4,m_5,m_6$ those of the three particles in the loop.
The coefficients $ h^I_{(T)}(m_1,\cdots,m_6)$ are in fact $(6j)$ symbols between simple
representations of $DSU(2)$ ($I=G$) or $ISU(2)$ ($I=0$).
These coefficients are computed in \cite{FNR} and, according to our convention, we have: 
\begin{eqnarray}\label{VG}
Y(m_1,m_2,m_3) \; h^G_{(T)}(m_i) & = & \frac{\pi}{8 \sin m_4 \; \sin m_5 \; \sin m_6} \frac{1}{\sqrt{D(m_i)}} \\
\text{with} \;\; 
D(m_i) & = & \left| 
\begin{array}{cccc}
1 & \cos m_1 & \cos m_2 & \cos m_3 \\
\cos m_1 & 1 & \cos m_6 & \cos m_5 \\
\cos m_2 & \cos m_6 & 1 & \cos m_4 \\
\cos m_3 & \cos m_5 & \cos m_4 & 1 
\end{array}
\right|
\end{eqnarray}
$D(m)$ is a Graam determinant. In the (physically interesting) case where all the masses $m_i$ are equal to the same value $m$, 
the Graam determinant simplifies and reads $D(m)=(1-\cos m)^3(3 \cos m+1)^{-1/2}$.

In the non-gravitational case, the evaluation of the same graph is given in term of the volume $V(m_1,\cdots,m_6)$ of a 
tetrahedron, whose length are fixed to the values $m_i$, by the formula:
\begin{eqnarray}\label{V0}
Y(m_1,m_2,m_3) \; h^0_{(T)}(m_i) \; = \; \frac{\pi}{48 \; m_1 \; m_2 \; m_3} \frac{1}{V(m_i)}\;.
\end{eqnarray}
When all the masses are fixed to the value $m$, the tetrahedron is regular and its volume is simply given by
$V(m)=\sqrt{2}m^3/12$.  Using the formula (\ref{VG})  and (\ref{V0}),  one obtains immediately that:
\begin{eqnarray}
h^G_{(T)}(m,\cdots,m) \; \equiv  \;h^G_{(T)}(m) =  \frac{1}{G^6} (1+\frac{13}{16}m^2G^2 + {\cal O}(G^3)) \; h^0_{(T)}(m) \;.
\end{eqnarray}
Then, quantum gravity corrections to the vertex one-loop corrections are immediately obtained from the previous expression 
(beware with the overall factor $G^{-6}$).

\vspace{0.3cm}

{\it c. Amplitudes and invariance under braidings}

\medskip

To precise what we mean by invariance of the amplitudes under braidings, we start by interpreting the amplitudes
computed previously as ``S matrix'' elements involving in and out states we have described in the section (2.2 ).

Let us focus on the case of the three-points function. On can naturally interpret the three-points functions as the deformed Fourier
transform of a certain S-matrix element:
\begin{eqnarray}
<\phi(\vec{x}_1) \phi(\vec{x}_2) \phi(\vec{x}_3)> \; = \; 
\int \prod_{k =1}^3 d\mu(\lambda_k) \; e^{i \text{tr}(\lambda_k h(m) \lambda_k^{-1} x_k)} \; 
<\lambda_2 \otimes \lambda_3 ; \ell_2 \otimes \ell_3 \; \vert \; \lambda_1 \otimes \ell_1> \;.
\end{eqnarray}
The S-matrix element $<\lambda_2 \otimes \lambda_3 ; \ell_2 \otimes \ell_3 \; \vert \; \lambda_1 \otimes \ell_1>$ gives the amplitude
between the in-state $\vert \lambda_1 \otimes \ell_1>$ and the out-state $\vert \lambda_2 \otimes \lambda_3 ; \ell_2 \otimes \ell_3 >$
when we have taken into account the gravitational interaction and the self-interaction. We have implicitely assumed that the observer
has zero mass ($m_0=0$); the field has a mass $m$. 

\begin{figure}[h]
\psfrag{x1}{$\vec{x}_1$}
\psfrag{x2}{$\vec{x}_3$}
\psfrag{x3}{$\vec{x}_2$}
\psfrag{x}{}
\psfrag{=}{$=$}
\psfrag{A1}{\!\!\!\!\!\!\!\!\!\!\!\!\!\!\!\!\!\!\!\!\!\!\!\!\!\!\!\!\!\!\!\!\!\!
$<\phi(\vec{x}_1) \tau(\phi(\vec{x}_2) \phi(\vec{x}_3))>_{(1)}=$}
\psfrag{A2}{\!\!\!\!\!\!\!\!\!\!\!\!\!\!\!\!\!\!\!\!\!\!\!\!\!\!\!\!\!\!\!\!\!\!
$<\phi(\vec{x}_1) \tau(\phi(\vec{x}_2) \phi(\vec{x}_3))>_{(3)}=$}
\centering
\includegraphics[scale=0.5]{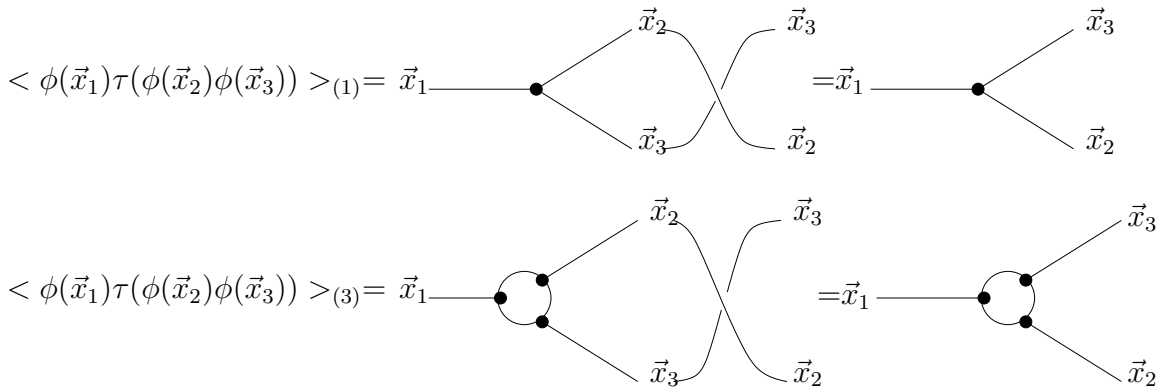}
\caption{Illustration of the invariance of the 3-points function under braidings: the first picture illustrates the invariance 
of the lowest order in $\lambda$ of the 3-points function; the second picture illustrates the case of a third order term. 
The crossings are a pictorial representation of the R-matrix. It is 
clear that the invariance is a consequence of the ``pivotal'' symmetry of the simple intertwiner represented by a big dot.}
\label{braidinvariance}
\end{figure}

The S-matrix element is, up to a factor ${\cal A}_G(m)$, an simple intertwining coefficient and therefore the three points function reads:
\begin{eqnarray}
<\phi(\vec{x}_1) \phi(\vec{x}_2) \phi(\vec{x}_3)> \; = \; {\cal A}_G(m) \;
\int \prod_{k =1}^3 d\mu(\lambda_k) \; e^{i \text{tr}(\lambda_k h(m) \lambda_k^{-1} x_k)} \; 
\delta(\prod_{k=1}^3 \lambda_k h(m)\lambda_k^{-1}) \; .
\end{eqnarray}
We have shown in the section (2.2) that the action of the braiding on a given particles-spin-network state 
reduces to an action of the R-matrix on the state and, as simple intertwiners are ``invariant'' under braiding, we have:
\begin{eqnarray}
<\phi(\vec{x}_1) \tau(\phi(\vec{x}_2) \phi(\vec{x}_3))> \; = \; <\phi(\vec{x}_1) \phi(\vec{x}_3) \phi(\vec{x}_2)> \;.
\end{eqnarray}
Therefore, the result of the braiding on the three-points function trivially reduces to the permutation of the arguments $\vec{x}_2$,
$\vec{x}_3$; the amplitude ${\cal A}_G(m)$ is unchanged. This property is illustrated in the figure (\ref{braidinvariance}).

In that sense, we claim that $n$-points functions are invariant under braidings.

\subsection*{4. Discussion and Generalization}
This article proposes a model for a three dimensional euclidean self-gravitating non-causal 
quantum field theory. The basic idea of our construction
is to quantize first gravitational degrees of freedom (using LQG techniques) before quantizing matter field degrees of freedom: 
$n$-particles states in a quantum background are defined as particles-spin-network states and form a physical Hilbert space;
the quantum self-gravitating field is described as usual in QFT as an operator acting on the self-gravitating Fock space, i.e. 
the infinite tower of $n$-particles physical Hilbert spaces. We focus only on the case of a massive spinless particles. The resulting
theory is a quantum field theory whose symmetry group is no longer the isometry group of the flat euclidean space $\mathbb E^3$ but
the quantum group $DSU(2)$ which can be viewed as a deformation of the classical group $ISU(2)$. In the lagrangian point of view, it
is clear that the theory is in fact a group field theory that can be easily written as a non-commutative quantum field theory 
(as it was first proposed by \cite{FreidelLivine}). The non-commutativity is a consequence of quantum gravity effects. 
Then, we generalize the model by introducing a self-interacting potential, whose coupling constant is $\lambda$,
that makes it more interesting. 

The nice feature with our model is that we can explicitely compute any terms of the series in the coupling $\lambda$ defining $n$-points
functions and then we can evaluate quantum gravity corrections. We have illustrated this property in the computation of lowest order
terms of the propagator (two-points function) and the vertex (three-points functions) of the self-gravitating self-interacting
quantum field theory. Lowest order (in the Newton constant $G$) quantum gravity corrections are explicitely computed. 

\medskip

Nevertheless, the model is not physical for it describes a three dimensional euclidean QFT. How to generalize to four dimensional case
is a complete open question even if LQG techniques work very well (at least at the kinematical level) even in four dimensions. 
How to construct a similar model with a lorentzian signature of space-time is a more suitable question. Making the theory
lorentzian is just a matter of technicity for one has to replace the quantum double $DSU(2)$ by its lorentzian counterpart.
It would be very nice to study and develop such a model for different reasons: recovering a good notion of causality, introducing a
time dimension and therefore defining a real dynamics for the quantum field. In the euclidean model, there is indeed no dynamics and there
is no canonical way to recover a certain dynamics. In fact, our article presents a canonical construction of a Group Field Theory that
reproduces non-causal spin-foam amplitudes of 3D gravity coupled to matter field. In that sense, one can say that our model is
a topological theory and does not behave as usual causal quantum field theory. In particular, the propagator is the gravitational analog
of the Hadamard propagator of the scalar field and not the usual Feynmann propagator. A natural way to recover the gravitational analog
of the Feynmann propagator is to add by hand a kinetic term to the action as it was done in \cite{FreidelLivine} or to impose a causality
relation at the level of the spin-foam model as it was done in \cite{OritiTlas}. The problem is that the resulting theory is apparently
non-unitary. This might come from the fact that there is no canonical way to define a causality relation for riemannian quantum field
theory. The behavior of the lorentzian model seems to be much nicer.
We hope to present the lorentzian model in great details in a future article.

\medskip

Even in the euclidean model, many points remain to be understood and deserve to be studied. First of all, we could generalize
the construction to the case where the space admits a non-trivial topology: in that case, the model would admit new types of degrees
of freedom and one could compute topology changing amplitudes in the presence of a matter field. Then, we could consider spinning
particles and could try to describe the self-gravitating quantum field theory. Therefore, we could in principle adapt our construction  
to describe a self-gravitating theory for fermions, gauge vectors and coupling theory involving different types of particles. We are
currently working in that direction.
Finally, we could also construct
the model in the presence of a cosmological constant: in that case, the quantum group structure would be changed into a more
interesting quantum group like $U_q(su(2))$ where the quantum deformation $q$ is related to the value of the cosmological constant.
We are currently working in that direction.

\subsubsection*{Aknowledgments}
I would like to thank J. Mourad and R. Parentani for interesting and stimulating discussions concerning quantum field theory.
My aknowledgments go to E. Livine as well for many interesting discussions.
Finally, I want to thank the referees for their suggestions to improve the presentation and the content of the paper.

{\it Et surtout... ma petite fille In\`es.}

\subsection*{Appendix A: Spin and Statistics for 3D euclidean QFT}
We review generalities concerning the relation between spin and statistics in quantum field theory in the first section of this appendix.
The second section illustrates the no spin-statistics relation in the special case of three dimensional euclidean quantum field
theory.
\subsubsection*{A.1. Generalities}
The classical configuration space ${\cal M}_n$ of a system of $n$ indistinguishable particles moving on a 
(connected and path connected) manifold locally $\mathbb R^d$ is given by:
\begin{eqnarray}
{\cal M}_n \; = \; (\mathbb R^{dn} -D)/S_n
\end{eqnarray}
where $S_n$ is the permutation group of $n$ elements and the ``diagonal'' $D$ is the set of singular configurations
where at least two particles coincide. 
\begin{enumerate}
\item Case $d=1$: ${\cal M}_n$ is clearly multiply connected (if $n>1$ of course)  and each of its connected component is topologically
equivalent to the sphere $S^{dn}$.
\item Case $d=2$: ${\cal M}_n$ is connected but not simply connected. Its fundamental group $\pi_1({\cal M}_n)$ is non trivial:
in fact $\pi_1({\cal M}_n)=B_n(\mathbb R^2)$, the braid group in dimension 2. $B_n(\mathbb R^2)$ is generated by $n-1$ elements
$\sigma_1,\cdots,\sigma_n$ which satisfy the following well-known algebraic relations:
\begin{eqnarray}
\sigma_i \sigma_j = \sigma_j \sigma_i \;\; \text{if} \;\; \vert i-j \vert \neq 1 \;\;,\;\;\;
\sigma_i \sigma_{i+1} \sigma_i  =  \sigma_{i+1} \sigma_{i} \sigma_{i+1} \;.
\end{eqnarray} 
\item Case $d>2$: ${\cal M}_n$ is connected; its fundamental group is the permutation group
$S_n$. Note that $S_n$ is generated by the $n-1$ elements $\sigma_i$ of $B_n(\mathbb R^2)$ satisfying the same relations as
the previous ones plus the condition $\sigma_i^2=1$.
\end{enumerate}
There is a consistent quantization associated to each unitary 
irreducible representations (irreps) of the fundamental group $\pi_1({\cal M}_n)$. Here, we are only interested with one dimensional
irreps even if higher dimensional irreps are mathematically conceivable. 
The statistics associated to these one dimensional irreps are the following (we will not consider the one dimensional case):
\begin{enumerate}
\item Case $d>2$: there exist two one dimensional representations $\chi_\pm$, the trivial one $\chi_+=1$ and the ``inverse'' one 
$\chi_-=-1$. The remarquable spin-statistics theorem claims that (upon some hypothesis we do not want to precise in details here)
fields which are symmetric under permutations (i.e. permutations are represented by $\chi_+$) are interger spin-fields (bosons) 
and fields which are anti-symmetric under permutations (i.e. permutations are represented by $\chi_-$) are half-integer spin-fields
(fermions). Therefore, there is a strong relation between representations of the Poincar\'e group and one-dimensional representations 
of $\pi_1({\cal M}_n)$.
\item Case $d=2$: one dimensional irreps $\chi_\theta$ are labelled by a real number $\theta \in [0,2\pi[$ such that 
$\chi_\theta(\sigma_i)=e^{i\theta}$ for any $i$. These statistics contains of course the bosonic ($\theta=0$) and the fermionic 
($\theta=\pi$) ones; the other statistics ($\theta \neq 0,\pi$) are known as anyonic statistics. There is generically 
no spin-statistics theorem in that case.
\end{enumerate}

\subsubsection*{A.2. 3D euclidean quantum field theory}
This section aims at illustrating the no spin-statistic \cite{SpinStatistic} theorem in a simple example of 3D euclidean quantum
field theory.

We start by considering massive spinless particles of mass $m$.
One-particle states on a three dimensional euclidean space
are represented by elements $\phi \in L^2(S^2,d\mu)$. For what follows, it is convenient to include in the space of states 
distributional states and then we introduce the bra-ket notation $\vert \lambda>$ to label a state of pure momentum 
($\lambda \in S^2 \subset SU(2)$). 
Such a state is 
defined by $<\phi \vert \lambda> = \phi(\lambda)$.
The no-particle state is denoted $\vert 0>$ and creation and annihilation operators $a^\dagger(\lambda)$ and $a(\lambda)$ acts as usual
on one and no-particle states. In particular,  we have
$a^\dagger(\lambda) \vert 0>=\vert \lambda >$.
Let us assume that these operators satisfy an anyonic statistic, i.e.:
\begin{eqnarray}
[a(\lambda_1);a^\dagger(\lambda_2)]_q & \equiv & a(\lambda_1) a^\dagger(\lambda_2) - q a^\dagger(\lambda_2)a(\lambda_1) \; = \;
\delta(\lambda_1^{-1} \lambda_2) \\
{}[a(\lambda_1);a(\lambda_2)]_q & = & 0 \; = \; [a^\dagger(\lambda_1);a^\dagger(\lambda_2)]_q \;\;.  
\end{eqnarray}
One can directly generalize these commutation relations for any $\lambda \in SU(2)$.
The statistic parameter $q$ is a priori any complex number. We want to see the compatibility between this type of statistics
and the value of the spin of the quantum field. To do so, we first need to define a spinning quantum field. 

One can make used of spinless particles creation/annihilation operators to construct spinning particles creation/annihilation
operators as follows:
\begin{eqnarray}
a^{\dagger}(\lambda,s) \; \equiv \; \int d\mu(\theta) \; e^{-i\theta s} \; a^{\dagger}(\lambda h(\theta)) \;\; , 
\;\;\;\; a(\lambda,s) \; \equiv \; \int d\mu(\theta) \; e^{+i\theta} \; a(\lambda h(\theta)) \;.
\end{eqnarray}
We have introduced the following notations: $s \in \frac{1}{2} \mathbb N$ is the spin of the particle; the group element 
$h(\theta)=\text{diag}(e^{i\theta},e^{-i\theta})$ in the $SU(2)$ spinorial representation and the measure is:
\begin{eqnarray}
d\mu(\theta) \; \equiv \; \lim_{n \rightarrow \infty} \int_{-n\pi}^{+n\pi} d\theta \;. 
\end{eqnarray}
It is easy to see that spinning particles creation/annihilation operators satisfy the following commutation relations:
\begin{eqnarray}
[a(\lambda_1,s_1);a^{\dagger}(\lambda_2,s_2)]_q \; = \; \delta(s_1 - s_2) \; \int d\mu(\theta) \; e^{-is \theta} 
\delta(\lambda_1^{-1} \lambda_2 h(\theta))\;.
\end{eqnarray}
A straightforward calculation shows that the state $\vert \lambda,s> \equiv a^\dagger(\lambda,s)\vert 0>$ is a (distributional) vector of 
a spinning massive representation of $ISU(2)$ and therefore represents a pure momentum spinning massive particle state. To be more
concrete, we have:
\begin{eqnarray}
\vert \lambda h(\alpha),s> \; = \; e^{i s \alpha} \; \vert \lambda,s> \;.
\end{eqnarray}
Then, a spinning quantum field operator $\phi_s(\vec{x})$ 
is defined as a linear combination of creation and annihilation operators; and the requirement of
locality and covariance imply that it takes the following form:
\begin{eqnarray}
\phi_s(\vec{x}) \; = \; \int d\mu(\lambda) \left(A(s) e^{-i\vec{x} \cdot \lambda \vec{m}} a^\dagger(\lambda,s)
+ B(s) e^{+i\vec{x} \cdot \lambda \vec{m}} a(\lambda,s) \right) \;.
\end{eqnarray}
$A(s)$ and $B(s)$ are complex coefficients that depend a priori of the spin $s$.
As our theory is euclidean, the causality requirement is replaced by the following conditions:
\begin{eqnarray}
[\phi_s(\vec{x});\phi_s(\vec{y})]_q & = & (1-q) \; A(s) B(s) \; \delta_s \;  \Delta_0(\vec{x} - \vec{y}) \; = \; 0\\
{}[\phi_s(\vec{x});\phi_s(\vec{y})^\dagger]_q & = & (\vert B(s) \vert^2 + q \vert A(s) \vert^2) \; \delta_s \;  
\Delta_0(\vec{x} - \vec{y}) \; = \; 0 \;.
\end{eqnarray}
The fonction $\Delta_0$ is the Haadamard propagator as defined in (\ref{ISU(2)2points}).
We see immediately that there is no obstruction to define a quantum spinning field whatever the statistics is: this is the illustration
of the no spin-statistics theorem in three dimensional space. 
There is only one subtelty concerning spinless particle: in that case, only the bosonic statistics ($q=-1$) is compatible with the
requirements of locality, covariance and causality. 

\subsection*{Appendix B: $DSU(2)$ as a gravitational deformation of $ISU(2)$}
The quantum double $DSU(2)$ has already been defined many times in the litterature
(see \cite{DPR} for the finite group case). This section aims at presenting $DSU(2)$
in such a way that it appears clearly as a quantum deformation of the Euclidean group $ISU(2)$. For clarity reasons, we will not enter
into mathematical details. The reader interested in such details is invited to go to \cite{BKM}.

\subsubsection*{B.1. The Euclidean group $ISU(2)$}
For that purpose, we start by recalling that $ISU(2)$ is the semi-product of $SU(2)$ by $\mathbb R^3$: its elements are usually denoted 
$(\vec{a},u) \in  \mathbb R^3 \times SU(2)$ and they satisfy the following algebra and (group like) co-algebra relations:
\begin{eqnarray}
(\vec{a}_1,u_1) \cdot (\vec{a}_2,u_2) \; = \; (u_1\vec{a}_2 + \vec{a}_1,u_1u_2) \;\;,\;\;\;\;
\Delta(\vec{a},u) \; = \; (\vec{a},u) \otimes (\vec{a},u).
\end{eqnarray}
The notation $u \vec{a}$ holds for the action of the vectorial representation of $u$ on the vector $\vec{a}$.
For latter convenience, it is useful to give the following equivalent description where the elements 
\begin{eqnarray}\label{classicalFT}
(\vec{k},u) \; \equiv \; \frac{1}{(2\pi)^3} \int d^3 \vec{a} \; e^{i\vec{k}\cdot \vec{a}} \; (\vec{a},u)
\end{eqnarray}
are ``Fourier'' transforms of the previous ones:
\begin{eqnarray}
(\vec{k}_1,u_1) \cdot (\vec{k}_2,u_2) \; = \; \delta^{(3)}(\vec{k}_1 - u_1 \vec{k}_2) \; (\vec{k}_1,u_1u_2) \;, \;\;
\Delta(\vec{k},u) \; = \; \int d^3\vec{p} \; (\vec{p},u) \otimes (\vec{k}-\vec{p},u) \;.
\end{eqnarray}

\subsubsection*{B.2. The Drinfeld double $DSU(2)$}
The quantum double $DSU(2)$ is defined as a coalgebra as the tensor product of $SU(2)$ with $Fun(SU(2))$ where $Fun(SU(2))$ is a suitable
set of functions on $SU(2)$. Its elements are usually denoted $(f,u) \in Fun(SU(2)) \times SU(2)$ in term of which the algebra and 
co-algebra structure are given by:
\begin{eqnarray}
(f_1,u_1) \cdot (f_2,u_2) \; = \; (f_1 \; f_2 \circ ad_{u_1},u_1u_2) \;, \;\;\;
\Delta(f,u) \; = \; \sum_{(f)}(f_{(1)},u) \otimes (f_{(2)},u) \;.
\end{eqnarray}
We have $(f\circ ad_u)(a)=f(u^{-1}au)$ and $(f_{(1)} \otimes f_{(2)})(a,b)=f(ab)$ for any $a,b \in SU(2)$.
As for the $ISU(2)$ case, it will be convenient in the sequel to deal with dual elements $(g,u) \in SU(2) \otimes SU(2)$
related to the previous one by:
\begin{eqnarray}
(f,u) \; \equiv \; \int dg \; f(g) \; (g,u)\;.
\end{eqnarray}
In terms of these elements, the Hopf algebra structure reads:
\begin{eqnarray}
(g_1,u_1) \cdot (g_2,u_2) \; = \; \delta(g_1^{-1}u_1g_2u_1^{-1}) \; (g_1,u_1u_2) \;,\;\;\;
\Delta(g,u) \; = \; \int dh \; (gh,u) \otimes (h^{-1},u)\;.
\end{eqnarray}

\subsubsection*{B.3. The relation between $ISU(2)$ and $DSU(2)$}
Using the notations introduced previously in this appendix, it is immediate to see that there exists a one parameter family
$\phi_G$ of algebra morphisms between $ISU(2)$ and $DSU(2)$ defined as follows:
\begin{eqnarray}
\phi_G \; : \; ISU(2) \; \longrightarrow \; DSU(2) \;\;\;\;\; (\vec{k},u) \; \longmapsto \; (e^{i G \vec{k} \cdot \vec{\sigma}},u)
\end{eqnarray}
where $\vec{\sigma}=(\sigma_1,\sigma_2,\sigma_3)$ are the Pauli matrices. 
In that sense, $DSU(2)$ is viewed as a gravitational deformation of $ISU(2)$ and $G$ is the Newton constant.
Obviously, $\psi_G$ is not a co-morphism; otherwise $ISU(2)$ and $DSU(2)$ would have been equivalent as Hopf algebra. 

\medskip

If $ISU(2)$ is the isometry group of the Euclidean space $\mathbb E^3$, $DSU(2)$ can be viewed as an isometry quantum group
of a deformed Euclidean space $\mathbb E^3_G = DSU(2)/SU(2)$. How to recover a good notion of position in $\mathbb E^3_G$?
A natural idea is to generalize the classical Fourrier transform introduced in (\ref{classicalFT}) to the deformed case in such a way that
the following diagram is commutative:
\begin{eqnarray}
(\vec{k},u) \in ISU(2) & \stackrel{FT_0}{\longmapsto} &
 (\vec{a},u) \; = \; \int d^3\vec{k} \; e^{i\vec{k}\cdot \vec{a}} \; (\vec{k},u) \nonumber \\
 \downarrow \phi_G & & \downarrow \phi_G \\
(g,u) \in DSU(2) &  \stackrel{FT_G}{\longmapsto} &
(\vec{a},u) \; = \; \int dg \; K(\vec{a},g) \; (g,u) \;. \nonumber
\end{eqnarray}
To make the diagram commutative, one can choose the kernel $K(\vec{a},g) \equiv e^{i \text{tr}(ga)}$ where we make the identification
$a=\vec{a} \cdot \vec{\sigma}$ and $\text{tr}$ is the trace in the fundamental $SU(2)$ (equivalently $su(2)$) representation. We recover 
naturally the same notion of position as in \cite{FreidelLivine}: $\vec{a}$ (in the last line of the diagram) can be viewed as
a position coordinates in $\mathbb E^3_G$.

To precise ``how much'' the space $\mathbb E^3_G$ is non-commutative, it is much more convenient to deal with the dual space 
$Fun(\mathbb E^3_G)$ with $Fun$ means a suitable set of functions. The product $\star$ between two such functions 
$f_1$ and $f_2$ is defined by Hopf duality as follows:
\begin{eqnarray}
(f_1 \star f_2)(\vec{a}) \equiv \sum_{(\vec{a})}f_1(\vec{a}_{(1)}) \;  f_2(\vec{a}_{(2)}) \;
\text{with} \;
\Delta(\vec{a}) = \sum_{(\vec{a})}\vec{a}_{(1)} \otimes \vec{a}_{(2)} = \int dg dh \; e^{iG \text{tr}(agh)} g \otimes h \;.
\end{eqnarray}
As a consequence, the product between two non-commutative plane waves of the type $w_g(\vec{a})\equiv e^{iG\text{tr}(g\vec{a})}$
is the same as the one found in \cite{FreidelLivine}:
\begin{eqnarray}
w_g \star w_h \; = \; w_{gh} \;.
\end{eqnarray}
Therefore, the non-commutativity is controled by the ``classical'' non-commutativity of the group $SU(2)$ itself.
In fact, this star-product is nothing but the convolution product on the group $SU(2)$. To see this point, we work with Fourier
transformed functions defined by:
\begin{eqnarray}
\tilde{f}: SU(2) \longrightarrow \mathbb C \;\;\;\;\;\; \tilde{f}(g) \; = \; \frac{1}{(2\pi)^3}\int d^3\vec{a} \; 
f(\vec{a}) \; e^{iG \text{tr}(ga)}
\end{eqnarray}
and we see immediately that:
\begin{eqnarray}
(\tilde{f}_1 \star \tilde{f}_2)(g) \; \equiv \; \frac{1}{(2\pi)^3}\int d^3\vec{a} \; 
(f_1 \star f_2)(\vec{a}) \; e^{iG \text{tr}(ga)} \; = \; (\tilde{f}_1 \circ \tilde{f}_2)(g)
\end{eqnarray}
where $\circ$ is the convolution product.

\subsection*{Appendix C: the three-points function}
This section is devoted to give the technical details of the computation of the classical and gravitational three-points
function.
\subsubsection*{C.1. Classical case}
A direct application of Feynmann rules leads to the following form of the free classical three-points function:
\begin{eqnarray}\label{H1}
H_{(V)}^0(\vec{x}_1,\vec{x}_2,\vec{x}_3) \;\equiv \; \int \prod_{\ell=1}^3 d^3\vec{p}_\ell\; 
\frac{\delta(p_\ell-m_\ell)}{P(m_\ell)} \; e^{i\vec{p}_\ell \cdot \vec{x}_\ell} \; \delta^{(3)}(\vec{p}_1+\vec{p}_2+\vec{p}_3)\;.
\end{eqnarray}
Replacing the function $P(m)$ by its expression and decomposing the delta function $\delta^{(3)}$ in plane waves, one obtains 
immediately that:
\begin{eqnarray}
H_{(V)}^0(\vec{x}_1,\vec{x}_2,\vec{x}_3) \;= \; \frac{2}{(4\pi)^4m_1^2m_2^2m_3^2} \int d^3\vec{x} \; 
\prod_{\ell =1}^3 {\Pi}_{\ell}(\vec{x}-\vec{x}_\ell)
\end{eqnarray}
with
\begin{eqnarray}
\Pi_\ell(\vec{x}) \; \equiv \; \int d^3\vec{p} \; \delta(\vert \! \vert \vec{p} \vert \! \vert - m_\ell) \; 
e^{i\vec{p}\cdot \vec{x}} 
\; = \; 4\pi m_\ell \; \frac{\sin(m_\ell \vert \! \vert \vec{x}  \vert \! \vert)}{ \vert \! \vert \vec{x}  \vert \! \vert} 
\end{eqnarray}
Therefore, we obtain the first expression for the classical three-points 
function:
\begin{eqnarray}
H^0_{(V)}(\vec{x}_1,\vec{x}_2,\vec{x}_3) \; = \; \frac{1}{(2\pi)^3}
\frac{1}{m_1 \; m_2 \; m_3} 
\int d^3\vec{x} \prod_{\ell=1}^3 
\frac{\sin(m_\ell \vert\!\vert \vec{x}-\vec{x}_\ell \vert \! \vert)}
{\vert\!\vert \vec{x}-\vec{x}_\ell \vert \! \vert} \;.
\end{eqnarray}
This expression is symmetric in the variables $\vec{x}_\ell$ but not very convenient. There exist other expressions we want to give here. Let us consider (\ref{H1}) and let us integrate over the variable $\vec{p}_3$. We write the remaining momenta in spherical coordinates, i.e. $\vec{p}_1=m_1\vec{n}_1$ and
$\vec{p}_2=m_2\vec{n}_2$ (after integrating over the delta functions involving the norms of the momenta), the expression (\ref{H1}) reduces to:
\begin{eqnarray}
H_{(V)}^0(\vec{x}_1,\vec{x}_2,\vec{x}_3) \; = \; \frac{1}{4\pi m_3^2}
\int d^2\vec{n}_1 \; d^2\vec{n}_2 \; \delta(p_{12} - m_3) 
e^{i(\vec{A}_1 \cdot \vec{n}_1 + \vec{A}_2 \cdot \vec{n}_2)}
\end{eqnarray}
where $\vec{A}_\ell = m_\ell(\vec{x}_\ell - \vec{x}_3)$ and
$p_{12} = \sqrt{\vert \! \vert m_1 \vec{n}_1 -  m_2 \vec{n}_2 \vert \! \vert}$
. Using the fact that:
\begin{eqnarray}
\delta(p_{12}-m_3) \; = \; 2m_3 \delta(p_{12}^2-m_3^2) \; = \;
\frac{m_3}{m_1m_2} \delta(\vec{n}_1\cdot \vec{n}_2 - M) \;\; \text{with} \;\;
M \; = \; \frac{m_1^2+m_2^2 - m_3^2}{2m_1m_2}
\end{eqnarray}
we obtain the following form:
\begin{eqnarray}
H_{(V)}^0(\vec{x}_1,\vec{x}_2,\vec{x}_3) \; = \; 
{\cal N}_0 \int d^2\vec{n}_1 \; d^2\vec{n}_2 \; 
\delta(\vec{n}_1 \cdot \vec{n}_2 - M) \; 
e^{i(\vec{A}_1\cdot \vec{n}_1 + \vec{A}_2\cdot \vec{n}_2)}.
\end{eqnarray}
${\cal N}_0^{-1}=4\pi m_1m_2m_3$. The integral vanishes if $\vert M \vert>1$;
in the non-vanishing case, we note $M=\cos S_0$ with $S_0$ a positive angle.
One can proceed as follows to integrate the last delta function in that expression: we decompose the vector $\vec{n}_2$ in the orthonormal basis 
$(\vec{n}_1,\vec{b},\vec{t})$ such that:
\begin{eqnarray}
\vec{n}_2 \; = \; \cos \theta \; \vec{n}_1 + 
\sin \theta(\cos \phi \; \vec{b} + \sin \phi \; \vec{t}) \;\;
\text{and} \;\; d^2\vec{n}_2 \; = \; \frac{1}{4\pi} \; \sin \theta \; d\theta d\phi;
\end{eqnarray}
and we obtain that:
\begin{eqnarray}
H_{(V)}^0(\vec{x}_1,\vec{x}_2,\vec{x}_3) \; = \;  
\frac{{\cal N}_0}{4\pi} \int d^2\vec{n}_1 && 
\int_0^\pi d\theta \; \sin \theta \; \delta(\cos \theta - \cos S_0) \nonumber \\
&& \int_0^{2\pi} d\phi \; 
e^{i(\vec{A}_1 + \cos \theta \vec{A}_2) \cdot \vec{n}_1}
e^{i\sin \theta(\cos \phi \vec{A}_2\cdot \vec{b} +
\sin \phi \vec{A}_2 \cdot \vec{t})}\;.
 \end{eqnarray}
As a consequence, one can integrate over the angular variable $\theta$, we
introduce the Bessel function of the first kind 
$J_0(z)=(2\pi)^{-1} \int_0^{2\pi} d\theta \; e^{iz\cos \theta}$ 
and the expression reduces to:
\begin{eqnarray}
H_{(V)}^0(\vec{x}_1,\vec{x}_2,\vec{x}_3) \; = \;  \frac{{\cal N}_0}{2}
\int d^2\vec{n} \; e^{i(\vec{A}_1 + \cos S_0 \vec{A}_2)\cdot \vec{n}}
J_0\left(\sin S_0 \sqrt{\vert \!\vert \vec{A}_2 \vert \!\vert^2 - 
(\vec{A}_2\cdot \vec{n}_2)^2}\right)\;.
\end{eqnarray}
We choose a system of coordinates where $\vec{A}_2 = A_2(1,0,0)$,
$\vec{A}_1 = A_1(\cos U,\sin U,0)$ and 
$\vec{n}=(\cos \theta,\sin\theta \cos \phi,\sin\theta \sin \phi)$
and we obtain the final expression:
\begin{eqnarray}
H_{(V)}^0(\vec{x}_1,\vec{x}_2,\vec{x}_3) = \frac{{\cal N}_0}{4}
\; \int_0^\pi \!\!d\theta \sin \theta 
e^{iA_1 \cos U \cos \theta} J_0(A_1 \sin U \sin \theta) \;  
e^{iA_2 \cos S \cos \theta} J_0(A_2 \sin S \sin \theta)  . 
\end{eqnarray}
Simplifications occur in some particular cases:
\begin{enumerate}
\item $S_0=0$: after some simple calculations, we see that:
\begin{eqnarray}
H_{(V)}^0(\vec{x}_1,\vec{x}_2,\vec{x}_3) = \frac{{\cal N}_0}{2} \int_0^1 dx \; \cos\left( (A_1 \cos U +A_2) x\right) 
J_0\left(A_1 \sin U \sqrt{1-x^2}\right)
\end{eqnarray}
Using a formula involving Bessel functions of the first kind, we simplify this expression as follows:
\begin{eqnarray}
H_{(V)}^0(\vec{x}_1,\vec{x}_2,\vec{x}_3) = \frac{{\cal N}_0}{2} \frac{\sin (\vert \! \vert \vec{A}_1 + \vec{A}_2 \vert \! \vert)}
{\vert \! \vert \vec{A}_1 + \vec{A}_2 \vert \! \vert} \;.
\end{eqnarray}
Furthermore, $S_0=0 \Longleftrightarrow m_3^2 = (m_1+m_2)^2$ and then we obtain the expression (\ref{Sim1}) 
given in the core of the article.
\item $U=0$: the same type of calculations leads in that case to the following expression:
\begin{eqnarray}
H_{(V)}^0(\vec{x}_1,\vec{x}_2,\vec{x}_3) = \frac{{\cal N}_0}{2} 
\frac{\sin \sqrt{(A_1+A_2 \cos U)^2 + A_2^2 \sin^2U}}{\sqrt{(A_1+A_2 \cos U)^2 + A_2^2 \sin^2U}}\;.
\end{eqnarray}
Using $\cos U =(m_3^2 - m_1^2 - m_2^2)/(2m_1m_2)$ we end up with the expression (\ref{Sim2}).
\end{enumerate}

Before going to the gravitational case, we present other equivalent formulations of the function $H_{(V)}^0$ that could be useful.
In fact, there is an identity involving Bessel functions and Gegenbauer polynoms $C_k^\lambda(t)$ defined for instance in \cite{ashrcroft}:
\begin{eqnarray}
e^{iA\cos \alpha \cos \beta} J_0(A \sin \alpha \sin \beta) \; = \; \sqrt{2 \pi} 
\sum_{k=0}^{+\infty} i^{-k} (\frac{1}{2} + k) \frac{J_{\frac{1}{2}+k}(A) C_k^{1/2}(\cos \alpha) C_k^{1/2}(\cos \beta)}
{\sqrt{A} C_k^{1/2}(1)}\;. 
\end{eqnarray}
$\Gamma$ is the well known Euler function and $J_n$ are Bessel functions. This identity allows to write $H^0_{(V)}$ as 
the following  indefinite series:
\begin{eqnarray}
H_{(V)}^0(\vec{x}_1,\vec{x}_2,\vec{x}_3) = \frac{\pi}{2 \sqrt{A_1 A_2}} \sum_{k=0}^\infty (\frac{1}{2}+k) J_{\frac{1}{2}+k}(A_1)
J_{\frac{1}{2}+k}(A_2) \frac{C_k^{1/2}(\cos S_0) C_k^{1/2}(\cos U)}{C_k^{1/2}(1)}\;.
\end{eqnarray}
This series is convergent.

\subsubsection*{C.2. Gravitational case}
In the gravitational case, the three-points function is defined by:
\begin{eqnarray}
H_{(V)}^G(\vec{x}_1,\vec{x}_2,\vec{x}_3) \; \equiv \; \frac{1}{2\pi^2} \int \prod_{\ell =1}^3 dg_\ell \; \delta_{m_\ell}(g_\ell)
e^{i\text{tr}(g_\ell x_\ell)} \; \delta(g_1g_2g_3) \;.
\end{eqnarray}
One can integrate out the variable $g_3$ for instance. We parametrize the two remaining variables as usual by:
$g_\ell  = \cos m \mathbb I  + \sin m  \vec{n}_\ell \cdot \vec{\sigma}$,  and we have:
\begin{eqnarray}
H_{(V)}^G(\vec{x}_1,\vec{x}_2,\vec{x}_3) \;  = \; \frac{1}{2\pi \sin m_3} \int d^2 \vec{n}_1 \; d^2\vec{n}_2 \; 
\delta(\text{tr}(g_1g_2) - 2\cos m_3) \; e^{i\text{tr}(x_1g_1+x_2g_2-x_3g_1g_2)}\;.
\end{eqnarray}
Furthermore,
\begin{eqnarray}
&&\text{tr}(g_\ell x_\ell)  =  \sin m_\ell \; \vec{n}_\ell \cdot \vec{x}_\ell \\
&&\text{tr}(g_1 g_2 x_3)  =  (\cos m_2 \sin m_1 \vec{n}_1 + \cos m_1 \sin m_2 \vec{n}_2 + 
\sin m_1 \sin m_2 \vec{n_1}\wedge \vec{n}_2)\cdot \vec{x}_3 \\
&&\delta(\text{tr}(g_1g_2) - 2\cos m_3)  =  \frac{1}{2 \sin m_1 \; \sin m_2} \; 
\delta(\vec{n}_1 \cdot \vec{n}_2 + \frac{\cos m_3 - \cos m_1 \cos m_2}{\sin m_1 \sin m_2})
\end{eqnarray}
and the previous expression becomes:
\begin{eqnarray}
H_{(V)}^G(\vec{x}_1,\vec{x}_2,\vec{x}_3) & = & {\cal N}_G 
\int d^2 \vec{n}_1 \; d^2\vec{n}_2 \delta(\vec{n}_1 \cdot \vec{n}_2 - \frac{\cos m_3 - \cos m_1 \cos m_2}{\sin m_1 \sin m_2})\nonumber \\
&&\exp[i(\sin m_1 \vec{n}_1 \cdot \vec{x}_1 + \sin m_2 \vec{n}_2 \cdot \vec{x}_2)] \\
&& \exp[-i(\cos m_2 \sin m_1 \vec{n}_1 + \cos m_1 \sin m_2 \vec{n}_2 + \sin m_1 \sin m_2 \vec{n}_1 \wedge \vec{n}_2)\cdot \vec{x}_3]\;.
\nonumber 
\end{eqnarray}
with ${\cal N}_G^{-1} = 4\pi \sin m_1 \sin m_2 \sin m_3$. After some trivial algebra, we recover the expression (\ref{expression2}) given
in the core of the article. To recover the first given expression (\ref{expression1}) of $H_{(V)}^0$, we just remark after 
some calculations that:
\begin{eqnarray}
&& \int d^3\vec{p}_3 \; \delta(\vert \! \vert \vec{p}_3 \vert \! \vert - \sin m_3) 
\delta^{(3)}(\vec{p}_3 + \cos m_2 \vec{p}_1 + \cos m_1 \vec{p}_2 + \vec{p}_1 \wedge \vec{p}_2) \\
&=& \delta(\vert \! \vert \cos m_2 \vec{p}_1 + \cos m_1 \vec{p}_2 + \vec{p}_1 \wedge \vec{p}_2 \vert \! \vert - \sin m_3) \nonumber \\
&=& 2\sin m_3 
\delta(\vert \! \vert \cos m_2 \vec{p}_1 + \cos m_1 \vec{p}_2 + \vec{p}_1 \wedge \vec{p}_2 \vert \! \vert^2 - \sin^2 m_3) \nonumber \\
&=& 2\sin m_3 \delta\left( (\vec{p}_1\cdot \vec{p}_2 - \cos m_1 \cos m_2 + \cos m_3)
(-\vec{p}_1\cdot \vec{p}_2 + \cos m_1 \cos m_2 + \cos m_3)\right) \nonumber \\
&=& 2 \sin m_3 \left[ \frac{\delta(\vec{p}_1\cdot \vec{p}_2 - \cos m_1 \cos m_2 + \cos m_3)}{2 \vert \cos m_3 \vert}
+ \frac{\delta(-\vec{p}_1\cdot \vec{p}_2 + \cos m_1 \cos m_2 + \cos m_3)}{2 \vert \cos m_3 \vert} \right]\;.\nonumber 
\end{eqnarray}
Therefore, the expressions (\ref{expression1}) and (\ref{expression2}) are not strickly speaking equivalent. In the particular case where
we identify $m_3$ and $(\pi - m_3)$, then the two delta functions in the last line of the previous calculations are equivalent
and the equality between the two expressions is true. We assume that for simplicity we do this identification. 
Therefore, the expression (\ref{expression1}) is proven.
To recover, the expression (\ref{VGJ}) of $H_{(V)}^0$, we proceed exactly in the same way as in the classical case. The simplifications
that occurs in the particular cases ${\cal S}_G=0$ or $U=0$ are easy to see and left to the reader.

\subsubsection*{C.3. First order quantum gravity corrections}
To compute first order quantum gravity corrections to the three-points function, we start with the following expression:
\begin{eqnarray}\label{zizou}
H_{(V)}^G(\vec{x}_1,\vec{x}_2,\vec{x}_3) \; = \; \frac{{\cal N}_G}{2} \int d^2\vec{n} \;
e^{i(\vec{B}_1 + \cos S_G \vec{B}_2)\cdot \vec{n}} J_0(\sin S_G R(\vec{n}))
\end{eqnarray}
where $\vec{B}_1$, $\vec{B}_2$ and $R(\vec{n})$ have been introduced in the core of the paper. To compute the classical limit, we
have to scale the length variables and the mass variables respectively by the Planck length and the Planck mass:
\begin{eqnarray}
\vec{x}_\ell \; \longmapsto \; \frac{\vec{x}_\ell}{\ell_p} \;\;\;\;\; m_\ell \; \longmapsto \; \frac{m_\ell}{m_p} \;\;\;\;
\text{with} \;\; \ell_p=G \;\; \text{and} \;\; m_p=\frac{1}{G}\;.
\end{eqnarray}
Then, we develop the integrand of the expression (\ref{zizou}) at the first order and the only contribution to first order is the 
following (all the others create contributions at least at the second order):
\begin{eqnarray}
R(\vec{n}) \; \simeq \; \sqrt{\vert \! \vert \vec{A}_2 \vert \! \vert^2 -(\vec{A}_2 \cdot \vec{n})^2 } \; - \; 
Gm_1m_2^2 \frac{\vec{x}_2 \wedge \vec{x}_3 \cdot \vec{n}}
{\sqrt{\vert \! \vert \vec{A}_2 \vert \! \vert^2 -(\vec{A}_2 \cdot \vec{n})^2 }}\;.
\end{eqnarray}
Furthermore, ${\cal N}_G=G^{-3}{\cal N}_0$; as a consequence, we have:
\begin{eqnarray}
H_{(V)}^0(\vec{x}_1,\vec{x}_2,\vec{x}_3) &\simeq & \frac{H_{(V)}^0}{G^{3}} + \frac{{\cal N}_0 m_1m_2^2}{G^2} 
\int d^2\vec{n} \; e^{i(\vec{A}_1 + \cos S_0 \vec{A}_2)\cdot \vec{n}} 
J_1\left(\sin S_0 \sqrt{\vert \! \vert \vec{A}_2 \vert \! \vert^2 -(\vec{A}_2 \cdot \vec{n})^2}\right) \nonumber \\
&& \;\;\;\;\;\;\;\;\;\;\;\;\;\;\;\;\;\;\;\;\;\;\;\;\;\;\;\;\;\;\;\;\;\;\;\;\;\;\;\;\;
\frac{\vec{x}_2 \wedge \vec{x}_3 \cdot \vec{n}}{\sqrt{\vert \! \vert \vec{A}_2 \vert \! \vert^2 -(\vec{A}_2 \cdot \vec{n})^2 }}
\end{eqnarray}
We used $J'_0(z)=-J_1(z)$. Then, the expression given in the core of the article for first order quantum gravity corrections 
(\ref{firstorderQGcor}) follows immediately.

\bibliographystyle{unsrt}

\end{document}